\documentclass[a4paper,11pt]{article}

\usepackage[numbers,sort&compress]{natbib}
\usepackage{jcappub} 

\usepackage[T1]{fontenc} 

\usepackage{latexsym}
\usepackage{graphicx}
\usepackage{amsmath}
\usepackage{amssymb}
\usepackage{amsfonts}
\usepackage{times}
\usepackage{xspace} 

\usepackage{subfigure}		
\usepackage[normalem]{ulem} 

\title{\boldmath Tailoring cosmologies in cubic shift-symmetric Horndeski gravity}

\author{Reginald Christian Bernardo}
\author{and Ian Vega}
\affiliation{National Institute of Physics, University of the Philippines \\ Diliman, Quezon City 1101, Philippines}

\emailAdd{rbernardo@nip.upd.edu.ph}
\emailAdd{ivega@nip.upd.edu.ph}

\abstract{We present a method for furnishing flat Friedman-Robertson-Walker spacetimes with nearly arbitrary dynamics in an important subclass of cubic Horndeski theory --  specifically shift-symmetric, cubic Horndeski theory with a vanishing conserved current. This builds on insight from previous work on the construction of static and spherically-symmetric hairy spacetimes in the same sector. The method is explicitly demonstrated by deriving exact analytical solutions describing an inflating universe and several power-law expansion scenarios, and by showing how the predicted evolution of the Hubble parameter in $\Lambda$CDM  can be fit to a particular choice of Horndeski model function. We fully characterize the classes of cosmological models that cannot be generated purely by selecting a Horndeski model function.}

\begin{document}
\maketitle
\flushbottom

\section{Introduction}
\label{sec:intro}

There is no shortage of experimental support in making the case for general relativity (GR) as the correct theory of gravity. Most recent among these is LIGO's historic observation of gravitational waves \cite{gw_1_ligo, gw_2_ligo, gw_3_ligo, gw_4_ligo, gw_170817_ligo}, which stands out not only for the technological breakthrough it represents, but also for its promise of a new and exciting observational window to the cosmos.  On much larger scales, the accuracy of the $\Lambda$CDM model in describing the evolution of the universe is also touted by many as another theoretical and experimental success of GR. However, the extrapolation of GR to all scales can remain deeply unsettling to some, for reasons that are well-known, not least of which is the lack of theoretical foundations for attendant elements such as dark matter and dark energy \cite{accelerating_universe_reiss, accelerating_universe_tonry, dark_energy_frieman, dark_energy_weinberg}, the tension between our theoretical expectations for the cosmological constant and its observed value \cite{weinberg_cc, carroll2001_cc, armendariz_cc}, and GR's famous incompatibility with quantum mechanics \cite{qg_deser1, qg_deser2, qg_hawking, qg_ashtekar, qg_martellini, gq_gomis, qg_donoghue, qg_boulware}. All these puzzles suggest that GR is far from giving us the final chapter in the ongoing story of gravity. 

The hope persists that one's fix for the problems of GR at small scales will also translate into an effective theory of alternative/modified gravity on galactic and cosmological scales with no need for dark matter and/or dark energy. But one of the major challenges of this framework is that, in going from very short to very long length scales, it has to be able to bypass solar-systems length scales where GR works extremely well. Certain screening mechanisms allow some alternative gravity theories to do precisely this \cite{screening_chakraborty, screening_brax, screening_schmidt, screening_davis,st_horndeski_vainshtein_dima, alternative_gravity_koyama, alternative_gravity_joyce}, and the study of alternative gravity theories and their phenomenology remains an area of active research.

The simplest gravity theory encompassing GR as an effective field theory is a scalar-tensor theory \cite{alternative_gravity_koyama, alternative_gravity_joyce, alternative_gravity_clifton, st_dark_energy_tsujikawa, horndeski_review_kobayashi2019, quintessence_review_tsujikawa, st_galileon_inflation_deffayet, kessence_seminal_armendariz, kessence_seminal_armendariz2, covariant_galileon_deffayet, st_horndeski_seminal, st_horndeski_revive_deffayet, st_honrdeski_revive_deffayet_2, st_horndeski_galileons_charmousis, beyond_Horndeski_gleyzes, beyond_Horndeski_gleyzes2, dhost_langlois, dhost_achour,dark_energy_dhost_crisostomi, st_egb_inflation_chakraborty, st_horndeski_cosmology_emond2018,st_galileon_inflation_kobayashi, st_galileon_inflation_kobayashi_2, st_galileon_inflation_burrage}. In this class of theories, a cosmologically-active dark energy degree of freedom is attached to a scalar field. Horndeski gravity, rediscovered as generalized galileons, is the most general scalar-tensor theory with second-order field equations, thus, evading Ostrogradski instabilities arising from the Hamiltonian being unbounded from below \cite{st_horndeski_seminal, st_horndeski_revive_deffayet, st_honrdeski_revive_deffayet_2, st_horndeski_galileons_charmousis}. Theories beyond Horndeski and DHOST theories are scalar-tensor theories with higher-order field equations but with a degeneracy that allows the evasion of Ostrogradsky instabilities. \cite{beyond_Horndeski_gleyzes, beyond_Horndeski_gleyzes2, dhost_langlois, dhost_achour}. 

Horndeski gravity is interesting for a number of reasons. For one, it carries a rather large parameter space that supports a broad phenomenology -- for instance, an inflationary scenario with no need of a cosmological constant \cite{dark_energy_dhost_crisostomi, st_egb_inflation_chakraborty, st_horndeski_cosmology_emond2018}. Some of its sectors also support hairy black holes \cite{st_no_hair_theorem_hui, st_no_hair_benkel, st_no_hair_theorem_sotiriou_1, st_no_hair_theorem_sotiriou_2,st_black_holes_sotiriou,st_horndeski_babichev, st_horndeski_solutions_babichev_2, st_horndeski_cosmological_tuning_babichev, st_horndeski_slow_rotation_bh_maselli, st_horndeski_neutron_stars_maselli, st_horndeski_solutions_kobayashi, st_horndeski_solutions_babichev_0, st_horndeski_solutions_rinaldi, st_horndeski_solutions_anabalon, st_horndeski_solutions_minamitsuji, st_horndeski_solutions_gaete}, and with them the potential for discriminators between gravity theories in the strong-field regime. Among the most widely studied non-GR subsets of Horndeski theory are quintessence, $k$-essence, kinetic braiding, covariant galileons, Brans-Dicke, Gauss-Bonnet gravity, and theories with derivative coupling to the Einstein tensor \cite{alternative_gravity_koyama, alternative_gravity_joyce, alternative_gravity_clifton, st_dark_energy_tsujikawa, horndeski_review_kobayashi2019, quintessence_review_tsujikawa, st_galileon_inflation_deffayet, kessence_seminal_armendariz, kessence_seminal_armendariz2, covariant_galileon_deffayet}. 

The neutron star collision marked by gravitational wave event GW170817 \cite{gw_170817_ligo} has dramatically changed the landscape of alternative gravity, and specifically, Horndeski theory. In our view, it has essentially wiped out Horndeski sectors whose tensor modes propagate differently from the speed of light \cite{dark_energy_creminelli, dark_energy_ezquiaga, st_horndeski_cosmology_baker, st_horndeski_cosmology_sakstein, st_horndeski_cosmology_bettoni,st_horndeski_copeland2018}. These are its quartic and quintic sectors, which include theories with derivative coupling to the Einstein tensor, Gauss-Bonnet theory, and the quartic and quintic covariant galileons. The quadratic and cubic Horndeski sectors that evade GW170817 constraints have thus become the focus of more intense scrutiny. Much recent effort has been devoted to exploring their cosmological predictions for the cosmic microwave background, baryon acoustic oscillation, integrated Sachs-Wolfe effect, and weak lensing \cite{st_horndeski_galileon_barreira_1, st_horndeski_galileon_barreira_2,st_horndeski_renk, st_horndeski_galileon_peirone,  horndeski_constraint_mancini2019, horndeski_constraint_noller2018, horndeski_constraint_komatsu2019}. These surviving quadratic and cubic sectors include quintessence, $k$-essence, cubic galileon, and kinetic gravity braiding.

The task at hand consists of placing further observational and theoretical constraints on quadratic and cubic model functions \cite{st_horndeski_cosmology_de_felice, horndeski_constraint_mancini2019, horndeski_constraint_noller2018, horndeski_constraint_komatsu2019,st_horndeski_kennedy2017,st_horndeski_kennedy2018,st_horndeski_kennedy2019}. 
One way to proceed is to use the dark energy effective field theory dictionary \cite{eft_cosmology_piazza, eft_de_gleyzes2, eft_de_cremineli, eft_de_gubitosi, eft_de_gleyzes,st_horndeski_kennedy2017,st_horndeski_kennedy2018,st_horndeski_kennedy2019} which translates Horndeski theories and beyond to cosmological effective field theory parameters which are directly tied to observations. However, in spite of these advances and the strong constraints set by GW170817, the vast majority of Horndeski theories still remain relatively unscathed.

In this paper, we present a method for constructing exact background cosmologies in cubic shift-symmetric Horndeski theories with a vanishing scalar current\footnote{While writing this manuscript, Ref. \cite{tailoring_cosmologies_fomin} appeared in the arXiv showing similar results. Specifically, it shows how GR's cosmological solutions associated with matter, radiation, and vacuum-dominated eras can be assigned to scalar-tensor Gauss-Bonnet gravity theories. Our work complements this by covering a different, and arguably more phenomenologically interesting, sector of Horndeski theory.}. This GW170817-surviving sector is particularly interesting because it houses hairy black holes and has recently been identified as No Run Gravity \cite{no_slip_linder, no_slip_cmb_brush, no_run_linder} -- a possible endgame for scalar-tensor theories if future cosmological data reveal no notable deviation from GR. With this method we can select nearly any cosmological dynamics and determine the unique cubic Horndeski model (with zero scalar current) that supports it. This is analogous to what is possible in quintessence models or GR plus a single scalar field in a potential. In contrast to quintessence though, where both the potential function, $V$, and its derivative enter the field equations, cubic shift-symmetric Horndeski with vanishing scalar current contains only a single free function, $G_X$, in its field equations. The background field equations of the latter are thus more tractable. As a result, whereas determining the quintessence potential for a given Hubble parameter evolution generally involves the numerical inversion of an integral, building the analogous cubic Horndeski theory involves solely algebraic steps. The simplicity of the field equations allows us to build compact, exact and analytical expressions for cubic Horndeski theories that admit typical cosmological evolutions such as inflation and power-law expansions. 

The rest of the paper proceeds as follows. In Section \ref{sec:horndeski_theory}, we set the stage with a brief overview of Horndeski theory and current constraints. In Section \ref{subsec:background}, we derive the field equations of shift-symmetric cubic Horndeski with a vanishing scalar current. Then we outline the method by first illustrating how an inflationary scenario can be tailored into a cubic Horndeski theory (Section \ref{subsec:inflation}) and derive the necessary conditions (Section \ref{subsec:necessary_conditions}) from which we formulate a recipe for assigning (almost) any given scale factor or Hubble parameter evolution to a cubic Horndeski theory (Section \ref{subsec:recipe}). As a direct application of the recipe, we present compact expressions for Horndeski theories describing power-law expanding universes (Section \ref{subsec:power_law}) and fit $\Lambda$CDM's predicted scale factor evolution to a cubic Horndeski theory (Section \ref{subsec:cubic_lcdm}). We also show that a nondynamical dark energy equation of state $w_\phi = -1$ is incompatible in cubic shift-symmetric Horndeski theory with vanishing scalar current (Section \ref{subsec:nondynamical_dark_energy}). We end the paper with a summary of our results and a discussion of possible future work.

We work with a mostly plus signature $(-,+,+,+)$ and adopt geometric units in which $G = c = 1$.

\section{Cubic Horndeski theory}
\label{sec:horndeski_theory}

Horndeski theory or generalized galileon theory \cite{alternative_gravity_clifton, alternative_gravity_joyce, alternative_gravity_koyama, st_horndeski_seminal} is described by the action
\begin{equation}
\label{eq:horndeski_theory}
S = \int d^4 x \sqrt{-g} \left[ L_2 + L_3 + L_4 + L_5 \right]
\end{equation}
where
\begin{eqnarray}
L_2 &=& G_2 \\
L_3 &=& -G_3 \Box \phi \\
L_4 &=& G_4 R + G_{4X} \left[ \left( \Box \phi  \right)^2 - \left( \nabla_\mu \nabla_\nu \phi \right)^2  \right] \\
L_5 &=& G_5 G_{\mu\nu} \left( \nabla^\mu \nabla^\nu \phi \right) - \frac{1}{6} G_{5X} \left[ \left( \Box \phi  \right)^3 - 3 \Box \phi \left( \nabla_\mu \nabla_\nu \phi \right)^2 + 2 \left( \nabla_\mu \nabla_\nu \phi \right)^3  \right]
\end{eqnarray}
where $G_i = G_i\left( \phi, X  \right)$ are arbitrary functions of the scalar field $\phi$ and its kinetic density $X = - g^{\alpha \beta} \left(\partial_\alpha \phi \right) \left( \partial_\beta \phi \right) / 2$. The shift-symmetric sector of the theory corresponds to the choice $G_i \left( \phi, X \right) = G_i \left( X \right)$ and, in this case, the equations of motion of the metric and scalar field can be written as
\begin{eqnarray}
\label{eq:covariant_einstein_equation} G_{\alpha \beta} &=& 8 \pi T_{\alpha \beta} \\
\label{eq:covariant_scalar_equation} \nabla_\alpha J^\alpha &=& 0
\end{eqnarray} 
where $T_{\alpha \beta}$ encodes the stress-energy tensor of both the matter and the scalar field sectors and $J^\alpha$ is the Noether current arising from the shift-symmetry $\phi \rightarrow \phi + \phi_0$ where $\phi_0$ is a constant. Being the most general scalar-tensor theory that evades Ostrogradski instabilities \cite{alternative_gravity_clifton, alternative_gravity_joyce, alternative_gravity_koyama}, Horndeski theory has been shown to accommodate a very rich phenomenology. Its covariant galileon limit is particularly interesting because it can account for cosmic acceleration without the need for a cosmological constant \cite{st_galileon_inflation_kobayashi, st_galileon_inflation_kobayashi_2, st_galileon_inflation_burrage}. Well-known scalar-tensor theories such as quintessence, $k$-essence, kinetic gravity braiding, Brans-Dicke theory, covariant galileons, Gauss-Bonnet theory, and GR are all special cases of Horndeski theory \cite{alternative_gravity_koyama, alternative_gravity_joyce, alternative_gravity_clifton, st_dark_energy_tsujikawa, horndeski_review_kobayashi2019, quintessence_review_tsujikawa, st_galileon_inflation_deffayet, kessence_seminal_armendariz, kessence_seminal_armendariz2, covariant_galileon_deffayet}.

Current constraints on Horndeski theory include LIGO's observation that the tensor speed excess, which measures the deviation of the speed of gravitational waves from that of light, vanishes with an uncertainty of one part in $10^{15}$ \cite{gw_170817_ligo}. This disfavors most of the quartic and quintic sectors (terms with $G_4$ and $G_5$) of the theory while leaving the quadratic and cubic sectors (terms with $G_2$ and $G_3$) unconstrained. The latter sectors, which include GR, quintessence and $k$-essence, kinetic gravity braiding, and Brans-Dicke sectors, all predict tensor perturbations that propagate at the speed of light \cite{dark_energy_creminelli, dark_energy_ezquiaga, st_horndeski_cosmology_baker, st_horndeski_cosmology_sakstein, st_horndeski_cosmology_bettoni,st_horndeski_copeland2018}. It must be noted that some work has shown the cubic galileon sector, $G_3 \sim X$, to be inconsistent with the cosmic microwave background, baryon acoustic oscillations, and the integrated-Sachs-Wolfe effect. There are similar claims that the covariant galileon model $\left( G_2 \sim G_3 \sim X, G_4 \sim G_5 \sim X^2 \right)$ is statistically ruled out by cosmological data including weak lensing \cite{st_horndeski_galileon_barreira_1, st_horndeski_galileon_barreira_2,st_horndeski_renk, st_horndeski_galileon_peirone}. Furthermore, recent studies have also shown that our best cosmological data on the cosmic microwave background and baryon acoustic oscillations cannot fully rule out Horndeski theory \cite{horndeski_constraint_mancini2019, horndeski_constraint_noller2018, horndeski_constraint_komatsu2019}. Nonetheless, due to the tightening of constraints in some Horndeski sectors, the community has increasingly narrowed its focus on the surviving quadratic and cubic sectors \cite{st_horndeski_kennedy2018,st_horndeski_kennedy2019, st_horndeski_cubic_appleby, st_horndeski_scaling_solutions_albuquerque, st_horndeski_solutions_minamitsuji_2, st_horndeski_qnm_tattersall, st_horndeski_hair_dressing_bernardo, no_slip_linder, no_slip_cmb_brush, no_run_linder}. For instance, kinetic gravity braiding was narrowed down as No Run Gravity \cite{no_slip_linder, no_slip_cmb_brush, no_run_linder} -- which might be scalar-tensor theories' final stand if future precision observations are able to pin the tensor speed excess, gravitational slip, and Planck mass run rate to zero. Stronger model-independent constraints are expected to come in the near future as more precision experiments reach their conclusion. 

For the rest of the paper, we shall focus on the Horndeski sector defined by the action
\begin{equation}
\label{eq:theory}
S = \int d^4 x \sqrt{-g} \left[ R + \left( X - 2 \Lambda \right) - G\left(X\right) \Box \phi  \right]
\end{equation}
or, equivalently, by
\begin{eqnarray}
\label{eq:power_G2} G_2 &=& X - 2 \Lambda, \\
\label{eq:power_G3} G_3 &=& G(X), \\
\label{eq:power_G4} G_4 &=& 1,
\end{eqnarray}
where $\Lambda$ is a cosmological constant and $G$ is an arbitrary function of the kinetic density that we refer to as the cubic model function. This is a subset of kinetic gravity braiding. We note here that we treat the cosmological constant $\Lambda$ in Eq.~(\ref{eq:theory}) as part of the scalar sector of the theory. Later on, we shall introduce another cosmological constant which enters the matter sector as a perfect fluid with negative pressure. It is the model dependence on the kinetic density, i.e. $G(X)$, that prevents cubic Horndeski from being physically equivalent to $k$-essence where there is an arbitrary model function $K$ instead for the quadratic Horndeski sector \cite{st_galileon_inflation_deffayet}. The scalar field's stress-energy tensor and the conserved scalar current arising due to shift symmetry can be shown to be 
\begin{equation}
\label{eq:stress_energy_scalar_covariant}
\begin{split}
8 \pi T_{\alpha \beta} =& \frac{1}{2} \left( X - 2 \Lambda \right) g_{\alpha \beta} + \frac{1}{2} \left( \partial_\alpha \phi \right) \left( \partial_\beta \phi \right) \\
& + \bigg[ -\frac{1}{2} G_{X} \Box \phi \left( \partial_\alpha \phi \right) \left( \partial_\beta \phi \right) + \frac{1}{2} G_{\mu} \left( \nabla^\mu \phi \right) g_{\alpha \beta} - G_{(\alpha} \left( \partial_{\beta )} \phi \right) \bigg] ,
\end{split}
\end{equation}
and 
\begin{equation}
\label{eq:scalar_current}
J^{\alpha} = - \left( \nabla^\alpha \phi \right) \left[ 1 - G_{X} \Box \phi \right] + G_{X} \left( \nabla^\alpha X \right),
\end{equation}
where $G_\alpha = \partial_\alpha G$ and $G_X = d G/ dX$. Not only is this theory phenomenologically interesting, it is also more analytically tractable compared to quintessence or single-scalar field potential models since there is only a single free function entering the field equations. In what follows, we will show that it is possible to fit any Hubble evolution into a cubic Horndeski theory with a vanishing scalar current. The vanishing of the scalar current is a common physical constraint imposed when obtaining black hole solutions and cosmological solutions in shift-symmetric Horndeski theory \cite{st_no_hair_theorem_hui, st_no_hair_benkel, st_no_hair_theorem_sotiriou_1, st_no_hair_theorem_sotiriou_2,st_black_holes_sotiriou,st_horndeski_babichev, st_horndeski_solutions_babichev_2, st_horndeski_cosmological_tuning_babichev, st_horndeski_slow_rotation_bh_maselli, st_horndeski_neutron_stars_maselli, st_horndeski_solutions_kobayashi, st_horndeski_solutions_babichev_0, st_horndeski_solutions_rinaldi, st_horndeski_solutions_anabalon, st_horndeski_solutions_minamitsuji, st_horndeski_solutions_gaete}. In black holes, this constraint arises from requiring regularity of the current, $J_\mu J^\mu = \left(J^r\right)^2 / f(r)$, particularly at the event horizon. Asymptotic flatness for galileon black holes then leads to the condition that $J^r = 0$ everywhere. In the context of cosmology, a vanishing scalar current is conveniently assumed in order to automatically satistfy the scalar field equation, as is done in by Babichev in Ref. \cite{st_horndeski_cosmological_tuning_babichev}. 

\section{Designer cosmologies in cubic Horndeski}
\label{sec:cosmological_solutions}

In this section, we lay down the main results of the paper. We begin by reviewing the standard field equations in Section \ref{subsec:background}. In the next section, we work on a concrete example of our approach by building an inflationary spacetime in a cubic Horndeski theory. In Section \ref{subsec:necessary_conditions}, we derive model-independent necessary conditions central to our cosmology-building strategy, which is subsequently detailed in Section \ref{subsec:recipe}. We then apply the recipe to determine cubic Horndeski theories that give rise to various cosmological dynamics, specifically power-law scale evolution (Section \ref{subsec:power_law}) and even $\Lambda$CDM (Section \ref{subsec:cubic_lcdm}). On a slightly different note, we use the necessary conditions to show that a nondynamical dark energy equation of state, $w_{\phi} = -1$, cannot result from a cubic Horndeski theory with vanishing scalar current (Section\ref{subsec:nondynamical_dark_energy}).

\subsection{Background field equations}
\label{subsec:background}

Consider the spatially-flat Friedmann-Robertson-Walker (FRW) background with the line element
\begin{equation}
\label{eq:line_element}
ds^2 = - dt^2 + a(t)^2 d\vec{x}^2
\end{equation}
where $a(t)$ is the scale factor. We shall work out the Einstein field equations with both the scalar field, $T_{\alpha \beta}$, and a perfect fluid matter source, $T^{(\text{M})}_{\alpha \beta}$:
\begin{equation}
\label{eq:einstein_equation_cosmology}
G_{\alpha \beta} = 8 \pi \left( T_{\alpha \beta} + T^{(\text{M})}_{\alpha \beta} \right) .
\end{equation}
The scalar field sector's stress-energy tensor for the general kinetic gravity braiding theory is given by \cite{st_horndeski_slow_rotation_bh_maselli}
\begin{equation}
\begin{split}
8 \pi T_{\alpha \beta}^{(\phi)} = & \frac{1}{2} G_2 g_{\alpha \beta} + \frac{1}{2} G_{2X} \left( \partial_\alpha \phi \right) \left( \partial_\beta \phi \right) \\
& + \bigg[ -\frac{1}{2} G_{3X} \Box \phi \left( \partial_\alpha \phi \right) \left( \partial_\beta \phi \right) + \frac{1}{2} G_{3\mu} \left( \partial^\mu \phi \right) g_{\alpha \beta} - G_{3(\alpha} \left( \partial_{\beta )} \phi \right) \bigg]
\end{split}
\end{equation}
and the perfect fluid stress-energy tensor is given by
\begin{equation}
T^{(\text{M})}_{\alpha \beta} = \left( \rho + P \right) u^\alpha u^\beta + P g^{\alpha \beta} 
\end{equation}
where $\rho$ and $P$ are the perfect fluid's energy density and pressure, respectively. We also assume that the scalar field possesses the same symmetry as the background:
\begin{equation}
\label{eq:scalar_field_spatially_uniform}
\phi = \phi(t) .
\end{equation}
As always, we consider the perfect fluid to be comoving with the background, i.e.
\begin{eqnarray}
u^\alpha &=& (1, 0, 0, 0) = \delta^\alpha_0 \\
u_\alpha &=& (-1, 0, 0, 0) = -\delta_\alpha^0 .
\end{eqnarray}
The metric and the stress-energy tensor of the perfect fluid can be written down as
\begin{equation}
g_{\alpha \beta} = - \delta_\alpha^0 \delta_\beta^0 + a(t)^2 \left( \delta_{\alpha \beta} - \delta_\alpha^0 \delta_\beta^0 \right)
\end{equation}
and
\begin{equation}
\label{eq:stress_energy_matter_frw}
T^{(\text{M})}_{\alpha \beta} = \rho \delta_\alpha^0 \delta_\beta^0 + P a^2 \left( \delta_{\alpha \beta} - \delta_\alpha^0 \delta_\beta^0 \right) ,
\end{equation}
respectively. Similarly, the Einstein tensor can be written as
\begin{equation}
\label{eq:einstein_tensor_frw}
G_{\alpha \beta} = 3 \left( \frac{\dot{a}}{a} \right)^2 \delta^0_\alpha \delta^0_\beta + \left( -\dot{a}^2 - 2 a \ddot{a} \right) \left( \delta_{\alpha \beta} - \delta^0_\alpha \delta^0_\beta \right)
\end{equation}
and the stress-energy tensor of the scalar field can be written as
\begin{equation}
\label{eq:stress_energy_scalar_frw}
\begin{split}
8\pi T_{\alpha \beta}^{(\phi)} = 
& \frac{1}{2} \left[ -G_2 + G_{2X} \dot{\phi}^2 + 3 G_{3X} \frac{\dot{a}}{a} \dot{\phi}^3 - G_{3\phi} \dot{\phi}^2 \right] \delta_\alpha^0 \delta_\beta^0 \\
& + \frac{a^2}{2} \left[ G_2 - \left( G_{3\phi} + G_{3X} \ddot{\phi} \right) \dot{\phi}^2 \right] \left( \delta_{\alpha \beta} - \delta_\alpha^0 \delta_\beta^0 \right) .
\end{split}
\end{equation}
The fact that the Einstein tensor and the stress-energy tensors for the scalar and matter fields have the same form as the metric is a consequence of the isotropy and homogeneity inherent in the background. By matching Eqs.~(\ref{eq:einstein_tensor_frw}), (\ref{eq:stress_energy_scalar_frw}), and (\ref{eq:stress_energy_matter_frw}) according to Eq. (\ref{eq:einstein_equation_cosmology}), we then obtain
\begin{equation}
3 \left( \frac{\dot{a}}{a} \right)^2 = 8 \pi \rho + \frac{1}{2} \left( -G_2 + G_{2X} \dot{\phi}^2 + 3 G_{3X} \frac{\dot{a}}{a} \dot{\phi}^3 - G_{3\phi} \dot{\phi}^2 \right)
\end{equation}
for the time component and
\begin{equation}
-\dot{a}^2 - 2 a \ddot{a} = 8 \pi P a^2 + \frac{a^2}{2} \left[ G_2 - \left( G_{3\phi} + G_{3X} \ddot{\phi} \right) \dot{\phi}^2 \right]
\end{equation}
for the spatial components. Inseting the Hubble parameter
\begin{equation}
H = \frac{\dot{a}}{a}
\end{equation}
in place of the scale factor $a$, we can equivalently write down the field equations as
\begin{eqnarray}
\label{eq:friedmann_first} 3H^2 &=& 8 \pi \rho_{\text{total}} \\
\label{eq:friedmann_almost} 2 \dot{H} + 3 H^2 &=& - 8 \pi P_{\text{total}} 
\end{eqnarray}
where 
\begin{eqnarray}
\rho_{\text{total}} &=& \rho + \rho_\phi \\
P_{\text{total}} &=& P + P_\phi \\
\label{eq:dark_energy_density_general} \rho_\phi &=& \frac{1}{16 \pi} \left( -G_2 +   G_{2X} \dot{\phi}^2 + 3 G_{3X} \frac{\dot{a}}{a} \dot{\phi}^3 - G_{3\phi} \dot{\phi}^2 \right) \\
\label{eq:dark_energy_pressure_general} P_\phi &=& \frac{1}{16\pi} \left[ G_2 - \left( G_{3\phi} + G_{3X} \ddot{\phi} \right) \dot{\phi}^2 \right] .
\end{eqnarray}
From the above expressions for the pressure and the energy density of the scalar, the dark energy equation of state can be written down as
\begin{equation}
\label{eq:w_de_kinetic_braiding}
w_\phi = -1 + \frac{ G_{2X} \dot{\phi}^2 - 2 G_{3\phi} \dot{\phi}^2 + \left( 3 H \dot{\phi} - \ddot{\phi} \right) \dot{\phi}^2 G_{3X} }{ - G_2 + G_{2X} \dot{\phi}^2 + 3 G_{3X} H \dot{\phi}^3 - G_{3\phi} \dot{\phi}^2 } .
\end{equation}
This expression agrees with Ref. \cite{st_dark_energy_tsujikawa} except for the sign of the cubic model function.

Using the field equations, we can obtain
\begin{equation}
\label{eq:friedmann_second}
\dot{H} + H^2 = - \frac{4\pi}{3} \left( \rho_{\text{total}} + 3 P_{\text{total}} \right).
\end{equation}
Eqs.~(\ref{eq:friedmann_first}) and (\ref{eq:friedmann_second}) are the first and second Friedmann equations. These equations look exactly like their GR counterparts except that there is a scalar field contribution to the energy density and pressure. As usual, the first and second Friedmann equations do not make a closed set of equations that can be readily integrated and one must first specify an equation of state that is usually parametrized as,
\begin{equation}
\label{eq:equation_of_state}
P = w \rho, 
\end{equation}
where $w$ is, in general, a time-dependent function. In GR, there are two field equations which can be integrated for the density $\rho$ and the Hubble parameter $H$. In Horndeski gravity, the scalar field also satisfies a field equation that couples its dynamical evolution with the metric. In this case, the Friedmann equations and the scalar field equation have to be integrated for the density $\rho$, the Hubble parameter $H$, and the scalar field $\phi$. In what follows, we consider the shift-symmetric sector of cubic Horndeski that is equipped with the scalar current \cite{st_horndeski_babichev}
\begin{equation}
\label{eq:scalar_current}
J^{\alpha}(x) = - \left( \partial^\alpha \phi \right) \left[ G_{2X} - G_{3X} \Box \phi \right] + G_{3X} \left( \partial^\alpha X \right) .
\end{equation}
In the FRW setting with a spatially-uniform scalar field, this scalar current reduces to
\begin{equation}
\label{eq:scalar_current_frw}
J^\alpha(x) = \delta^\alpha_0 \dot{\phi} \left[ G_{2X} + 3 G_{3X} H \dot{\phi} \right].
\end{equation}
There is only a time component because of the symmetries of the background and the scalar field. The solutions presented in this paper and their Horndeski theories are restricted to the space of solutions with vanishing scalar current \cite{st_horndeski_cosmological_tuning_babichev}. Such a restriction is not necessary but is required for one to evade the no-hair theorem for galileons and thus to accommodate hairy black holes \cite{st_no_hair_theorem_hui, st_no_hair_benkel, st_no_hair_theorem_sotiriou_1, st_no_hair_theorem_sotiriou_2,st_black_holes_sotiriou,st_horndeski_babichev, st_horndeski_solutions_babichev_2, st_horndeski_cosmological_tuning_babichev, st_horndeski_slow_rotation_bh_maselli, st_horndeski_neutron_stars_maselli, st_horndeski_solutions_kobayashi, st_horndeski_solutions_babichev_0, st_horndeski_solutions_rinaldi, st_horndeski_solutions_anabalon, st_horndeski_solutions_minamitsuji, st_horndeski_solutions_gaete}. For the cosmological context we are presently concerned with, the vanishing of scalar current given by Eq. (\ref{eq:scalar_current_frw}) implies that either $\dot{\phi} = 0$ or $G_{2X} + 3 G_{3X} H \dot{\phi} = 0$. The former option just recovers GR, since a constant scalar field has a vanishing stress-energy tensor. We shall not be concerned with this branch, even though it might still be interesting if one studies perturbations. For the rest of this paper, we shall focus our attention to the other branch where the cosmological spacetime might be equipped with a nontrivial scalar field. This non-GR branch also contains the hairy black holes in shift-symmetric Horndeski theory \cite{st_no_hair_theorem_hui, st_no_hair_benkel, st_no_hair_theorem_sotiriou_1, st_no_hair_theorem_sotiriou_2,st_black_holes_sotiriou,st_horndeski_babichev, st_horndeski_solutions_babichev_2, st_horndeski_cosmological_tuning_babichev, st_horndeski_slow_rotation_bh_maselli, st_horndeski_neutron_stars_maselli, st_horndeski_solutions_kobayashi, st_horndeski_solutions_babichev_0, st_horndeski_solutions_rinaldi, st_horndeski_solutions_anabalon, st_horndeski_solutions_minamitsuji, st_horndeski_solutions_gaete}.

Setting $G_{2X} + 3 G_{3X} H \dot{\phi} = 0$, the field equations reduce to
\begin{eqnarray}
\label{eq:friedmann_1_cubic} 3 H^2 &=& 8 \pi \rho + \frac{1}{2} \left( \frac{ \dot{\phi}^2 }{2} + 2 \Lambda + 3 G_{X} H \dot{ \phi }^3 \right) \\
\label{eq:friedmann_2_cubic} \dot{H} + H^2 &=& - \frac{4\pi}{3} \rho \left( 1 + 3 w \right) - \frac{ \dot{\phi}^2 }{6} \\
& & + \frac{\Lambda}{3}  + \frac{1}{4} G_{X} \dot{\phi}^2 \left( \ddot{\phi} - H \dot{\phi} \right) \nonumber \\
\label{eq:scalar_current_vanishes} 1 + 3 G_{X} H \dot{\phi} &=& 0 .
\end{eqnarray}
In what follows, we shall manipulate these field equations, first by using Eq. (\ref{eq:scalar_current_vanishes}) to remove the model dependences, represented by $G_X$, in the Friedmann equations. One of the model-independent Friedmann equations will then be used to remove one of the three variables, $(H, \rho, \phi)$, that appears in the other equations. For static and spherically-symmetric solutions, this sequence of steps left us with an equation that also holds in GR \cite{st_horndeski_hair_dressing_bernardo}. This also happens in the present cosmological context, where we shall be led to what corresponds to the conservation of energy. 

\subsection{An inflationary solution}
\label{subsec:inflation}

Before laying down the recipe, let us first carry out these steps explicitly for the concrete de Sitter case, where the Hubble parameter $H$ has the constant value $H_i$ or, alternatively,
\begin{equation}
\label{eq:alternative_inflation_de_sitter}
a(t) = a_i \exp(H_i t) .
\end{equation}
In the de Sitter limit the field equations become
\begin{eqnarray}
\label{eq:ds_friedmann_first} 3 H_i^2 &=& 8 \pi \rho + \frac{1}{2} \left( \frac{ \dot{\phi}^2 }{2} + 2 \Lambda + 3 G_{X} H_i \dot{ \phi }^3 \right) \\
\label{eq:ds_friedmann_second} H_i^2 &=& - \frac{4\pi}{3} \rho \left( 1 + 3 w \right) - \frac{ \dot{\phi}^2 }{6} \\
& & + \frac{\Lambda}{3}  + \frac{1}{4} G_{X} \dot{\phi}^2 \left( \ddot{\phi} - H_i \dot{\phi} \right) \\
\label{eq:ds_scalar_field_equation} 1 + 3 G_{X} H_i \dot{\phi} &=& 0 .
\end{eqnarray}
With the aid of Eq. (\ref{eq:ds_scalar_field_equation}) to remove $G_X$ in Eqs. (\ref{eq:ds_friedmann_first}) and (\ref{eq:ds_friedmann_second}) we then get
\begin{eqnarray}
\label{eq:EE_1_alt_inflation} 3 H_i^2 &=& 8 \pi \rho + \frac{1}{2} \left( - \frac{ \dot{\phi}^2 }{2} + 2\Lambda \right) \\
\label{eq:EE_2_alt_inflation} H_i^2 &=& - \frac{4\pi}{3} \rho \left( 1 + 3 w \right) + \frac{1}{6} \left[ - \frac{\dot{\phi}^2}{2} + 2\Lambda \right] - \frac{\dot{\phi}\ddot{\phi}}{12 H_i} .
\end{eqnarray}
Eliminating the density $\rho$ gives 
\begin{equation}
\dot{\phi} \ddot{\phi} + \frac{3}{2} H_i \left( 1 + w \right) \dot{\phi}^2 = - 3 H_i \left( 1 + w \right) \left( 6 H_i^2 - 2\Lambda \right)
\end{equation}
which is a first-order differential equation for $\dot{\phi}$. The solution to this is given by
\begin{equation}
\label{eq:alternative_inflation_kinetic_density}
\dot{ \phi }^2 = e^{-3 H_i t \left( 1 + w \right) + 2K } - 12 H_i^2 + 4 \Lambda 
\end{equation}
where $K$ is an integration constant. It is easy to verify that this also satisfies the field equations given by Eqs. (\ref{eq:EE_1_alt_inflation}) and (\ref{eq:EE_2_alt_inflation}), valid in the space where the time component of the scalar current vanishes, with the perfect fluid energy density
\begin{equation}
\label{eq:alternative_inflation_matter_density}
\rho = \frac{1}{32 \pi}  e^{-3 H_i t \left( 1 + w \right) + 2K } .
\end{equation}
Having been derived from model-independent necessary conditions, the solutions presented here for $\phi$ and $\rho$ are, respectively, the only possible scalar field and perfect fluid energy density compatible with the de Sitter expansion. To associate the solutions with a cubic Horndeski theory requires Eq. (\ref{eq:ds_scalar_field_equation}) which, in this case, leads to
\begin{equation}
G_X = - \frac{1}{3 H_i \sqrt{2X} }
\end{equation}
or equivalently
\begin{equation}
\label{eq:alternative_inflation_model}
G(X) = - \frac{ \sqrt{2X} }{3 H_i} + C
\end{equation}
where $C$ is another integration constant. The addition of the integration constant $C$ is for completeness since a constant $G_3$ term in the Horndeski action can be eliminated using integration by parts. It is easy to verify that the original field equations are satisfied by the solutions presented above.

It is useful to note that, apart from imposing that $w$ be a constant, we did not have to specify the component of the perfect fluid. The inflationary scenario presented by Eqs. (\ref{eq:alternative_inflation_de_sitter}), (\ref{eq:alternative_inflation_kinetic_density}), (\ref{eq:alternative_inflation_matter_density}), and (\ref{eq:alternative_inflation_model}) holds for any matter perfect fluid equation of state and contains two arbitrary integration constants. An interesting and useful feature of the theory given by Eq. (\ref{eq:alternative_inflation_model}) is that the model function is independent of the equation of state and the integration constant $K$ \footnote{For static and spherically-symmetric spacetimes \cite{st_horndeski_hair_dressing_bernardo}, the method usually leads to integration constants that have to be fixed later on with the theory parameters.}. This makes it possible for matter or radiation, or any fluid component, to drive the inflationary scenario given by Eq. (\ref{eq:alternative_inflation_de_sitter}). The scalar field or the dark energy density is given by
\begin{equation}
\rho_\phi = \frac{1}{16\pi} \left( 6 H_i^2 - \frac{1}{2} e^{ -3 H_i t \left( 1 + w \right) + 2K } \right) 
\end{equation}
and the total energy density is 
\begin{equation}
\rho_{\text{total}} = \frac{3 H_i^2}{8\pi} .
\end{equation}
For baryonic and radiative material, the perfect fluid and scalar field energy densities decline exponentially while the total energy density is kept constant. When the perfect fluid acts as a cosmological constant, matter with negative pressure and equation of state $w = -1$, we obtain the same picture as in GR: the matter and scalar field energy densities are both constant in time so naturally the total energy density is a constant.

The dark energy equation of state can be shown to be
\begin{equation}
\label{eq:dark_energy_w_inflation}
w_\phi = -1 + \frac{ 1 + w }{ 1 - 12 H_i^2 e^{3 H_i t \left( 1 + w \right) - 2K} } .
\end{equation}
The dark energy equation of state $w_\phi$ is constrained to be very close to $-1$. We see from Eq. (\ref{eq:dark_energy_w_inflation}) that if the perfect fluid component is a cosmological constant, $w = -1$, then the dark energy equation of state is the same as the perfect fluid. In this case, the scalar field's kinetic energy and energy densities reduce to a constant and are indistinguishable from the matter sector's cosmological constant. The scalar field sector is then absorbed into the matter sector and the theory simply reduces to GR. We can also use Eq. (\ref{eq:dark_energy_w_inflation}) to see what other fluid components can source an observed accelerated expansion. Other than using a cosmological constant fluid to get the dark energy equation of state $w_\phi$ close to $-1$, the theory allows us to choose the integration constant $K$ to be a very large negative number so as to make the second term in Eq. (\ref{eq:dark_energy_w_inflation}) negligible compared to unity. In this limit, the perfect fluid energy density becomes exponentially suppressed, $\rho \sim e^{-2 |K|}$, while the dark energy density approaches the expansion's energy, $\rho_\phi \sim 3 H_i^2/ 8\pi$. The other limit where the integration constant $K$ approaches a large positive value leads to $w_\phi \sim w$ while the matter and dark energy densities exponentially blow up while keeping $\rho_\phi / \rho \sim -1$.

\subsection{Necessary conditions for cosmological solutions}
\label{subsec:necessary_conditions}

We can generalize the foregoing calculation. In this section and the next, we derive necessary conditions for cosmological solutions arising from the field equations, and with these formulate a recipe for engineering cosmological solutions in cubic Horndeski. As already hinted at in the previous section, our first step is to use Eq. (\ref{eq:scalar_current_vanishes}) to eliminate $G_X$ in the Friedmann equations to obtain
\begin{eqnarray}
\label{eq:friedmann_1_model_independent} 3 H^2 &=& 8 \pi \rho + \frac{1}{2} \left( - \frac{\dot{\phi}^2}{2} + 2 \Lambda \right) \\
\label{eq:friedmann_2_model_independent} \dot{H} + H^2 &=& - \frac{4\pi}{3} \rho \left( 1 + 3 w \right) + \frac{\Lambda}{3} - \frac{\dot{\phi}^2}{12} - \frac{\dot{\phi} \ddot{\phi}}{12 H}.
\end{eqnarray}
These are necessary conditions demanded by the Friedmann equations in the space of solutions with a vanishing scalar current. Eq. (\ref{eq:friedmann_1_model_independent}) supplies us with an expression for the kinetic density in terms of the Hubble parameter:
\begin{equation}
\label{eq:necessary_X_cosmological}
X = 16 \pi \rho - 6 H^2 + 2\Lambda .
\end{equation}
In building the theory compatible with a given Hubble parameter, $H(t)$, this is the expression that we have to invert to $t(X)$ to write down $H(X)$ and obtain the model function from Eq. (\ref{eq:scalar_current_vanishes}). But before this, we need to know $H(t)$ and $\rho(t)$ because we want to write down the kinetic density $X$ explicitly as a function of the time coordinate $t$. Using Eq. (\ref{eq:friedmann_1_model_independent}) to eliminate $\phi$ in Eq. (\ref{eq:friedmann_2_model_independent}) we obtain
\begin{equation}
\label{eq:conservation_of_energy_cosmological}
\frac{\dot{\rho}}{\rho} + 3 \left( 1 + w \right) H = 0 .
\end{equation} 
The appearance of Eq. (\ref{eq:conservation_of_energy_cosmological}), which is just the statement of conservation of energy of the perfect fluid in GR, is both surprising and not surprising. For one, a GR-valid constraint also results when formulating the analogous recipe for dressing static and spherically-symmetric spacetimes with scalar hair \cite{st_horndeski_hair_dressing_bernardo}. It may also be argued to be unsurprising though considering that we are working in the Jordan frame where there is no direct coupling between the scalar field and matter fields. Thus, matter follows the same path as it would in GR with the same background. Other than this we can question why it is not the conservation of total stress-energy, which is guaranteed by one of the Bianchi identity, that appears instead of conservation of matter stress-energy only. The answer seems to be that the scalar and matter stress-energy tensors are separately conserved in the Horndeski sector that we are working in so that the conservation of matter stress-energy is equivalent to the conservation of total stress-energy. 

Recall that the Bianchi identity guarantees conservation of the total stress-energy tensor:
\begin{equation}
\label{eq:conservation_stress_energy_total}
\nabla_\beta \left( T^{(\phi) \alpha \beta} + T^{(\text{M}) \alpha \beta } \right) = 0,
\end{equation}
where $T^{(\phi) \alpha \beta}$ and $T^{(\text{M}) \alpha \beta }$ are the scalar and matter sectors stress-energy tensors, respectively. The fluid is, of course, subject to dynamics of its own so that we have 
\begin{equation}
\label{eq:conservation_stress_energy_matter}
\nabla_\beta T^{(\text{M}) \alpha \beta } = 0
\end{equation}
and a conservation of mass statement $\nabla_\beta \left( \rho_0 u^{\alpha} \right) = 0$ where $\rho_0$ stands for the mass density of the fluid. The conservation of energy statement given by Eq. (\ref{eq:conservation_of_energy_cosmological}) is, of course, nothing by the time component of Eq. (\ref{eq:conservation_stress_energy_matter}). It follows from Eq. (\ref{eq:conservation_stress_energy_total}) that the stress-energy tensor of the scalar is separately conserved from matter. Thus, we also have
\begin{equation}
\label{eq:conservation_stress_energy_scalar}
\nabla_\beta T^{( \phi ) \alpha \beta } = 0 .
\end{equation}
The interaction of the matter and the scalar field is only indirectly through the metric tensor. We'll explicitly show that the time component of Eq. (\ref{eq:conservation_stress_energy_scalar}) is satisfied. First, we note that 
\begin{equation}
\label{eq:stress_energy_scalar_frw_up}
\begin{split}
8\pi T^{(\phi) \alpha \beta} = 
& \frac{1}{2} \left[ -G_2 + G_{2X} \dot{\phi}^2 + 3 G_{3X} \frac{\dot{a}}{a} \dot{\phi}^3 \right] \delta^\alpha_0 \delta^\beta_0 \\
& + \frac{1}{2 a^2} \left[ G_2 -  G_{3X} \ddot{\phi} \dot{\phi}^2 \right] \left( \delta^{\alpha \beta} - \delta^\alpha_0 \delta^\beta_0 \right) .
\end{split}
\end{equation}
Also, the time component of the conservation of stress-energy, $\nabla_\beta T^{( \phi ) \alpha \beta } = 0$, can be written as
\begin{equation}
\dot{T^{00}} + 3 H T^{00} + \sum_i \left( \Gamma^t_{ii} T^{ii} \right) = 0 .
\end{equation}
In the space of solutions satisfying Eq. (\ref{eq:scalar_current_vanishes}), it can be shown that
\begin{eqnarray}
8\pi T^{(\phi) 00} &=& -\frac{\dot{\phi}^2}{4} + \Lambda \\
8\pi \dot{ T }^{(\phi) 00} &=& - \frac{1}{2} \dot{ \phi } \ddot{\phi} \\
8\pi \sum_i \left( \Gamma^t_{ii} T^{ii} \right) &=& 3 \left[ \frac{\dot{\phi} \ddot{\phi} }{6} + \left( - \Lambda + \frac{\dot{\phi}^2}{4} \right) H \right] .
\end{eqnarray}
Finally,
\begin{equation}
\begin{split}
\dot{T^{00}} + 3 H T^{00} + \sum_i \left( \Gamma^t_{ii} T^{ii} \right) = &- \frac{1}{2} \dot{ \phi } \ddot{\phi} + 3 H \left( -\frac{\dot{\phi}^2}{4} + \Lambda \right)  + 3 \left[ \frac{\dot{\phi} \ddot{\phi} }{6} + \left( - \Lambda + \frac{\dot{\phi}^2}{4} \right) H \right] .
\end{split}
\end{equation}
The right hand side vanishes exactly as expected, thus, we have
\begin{equation}
\nabla_\beta T^{( \phi ) 0 \beta } = 0 . 
\end{equation}
This explains the appearance of Eq. (\ref{eq:conservation_of_energy_cosmological}).

\subsection{Recipe for cosmological solutions}
\label{subsec:recipe}

In summary, the necessary conditions for building cosmological solutions are
\begin{eqnarray}
\label{eq:necessary_conserve_energy} \frac{\dot{\rho}}{\rho}\left[H\right] &=& - 3 \left( 1 + w \right) H   \\
\label{eq:necessary_X} X \left[H\right] &=& 16 \pi \rho \left[H\right] - 6 H^2 + 2 \Lambda \\
\label{eq:current_constraint} G_X\left[H\right] &=& -1/ 3H \dot{\phi}\left[H\right].
\end{eqnarray}
The strategy for building exact cosmological scenarios is analogous to what we previously laid out for dressing static and spherically-symmetric spacetimes with scalar hair in the same shift-symmetric cubic Horndeski sector \cite{st_horndeski_hair_dressing_bernardo}. 

First, we choose a cosmological evolution, $H(t)$ (e.g. for inflation, $H(t)$ is a constant), and then solve Eq. (\ref{eq:necessary_conserve_energy}) for the perfect fluid energy density. Having $H(t)$ and $\rho(t)$ then allows us to build the kinetic density $X(t)$ using Eq. (\ref{eq:necessary_X}). The kinetic density is the only other contribution to the dark energy density of the theory apart from the cosmological constant $\Lambda$. The remaining condition, Eq. (\ref{eq:current_constraint}), can be put to good use by exploiting the shift-symmetric restriction that the model function $G_X$ depends only on the kinetic density of the scalar field. This is the essential insight that allows us to generate cubic Horndeski theories with static and spherically-symmetric spacetimes equipped with nontrivial scalar field profiles. After inverting the expression $X(t)$, obtained from the previous step, we then use Eq. (\ref{eq:current_constraint}) to build the only cubic Horndeski theory that accepts the cosmological scenario specified by $H(t)$, the matter density $\rho(t)$, and the scalar hair $X(t)$. It can be verified that the solutions built in this way, exploiting model-independent necessary conditions, indeed satisfy the model-dependent Friedmann equations (Eqs. (\ref{eq:friedmann_1_cubic}) and (\ref{eq:friedmann_2_cubic})). The scalar field equation is naturally satisfied since we are working in the sector of Horndeski theory with vanishing scalar current.

If the cosmological scenario describes an accelerating universe, the dark energy equation of state is given by
\begin{equation}
\label{eq:w_de_H}
w_\phi = \frac{P_\phi}{\rho_\phi} = -1 - \frac{1}{3} \frac{ \dot{\rho}_\phi / \rho_\phi }{ H } 
\end{equation}
where
\begin{eqnarray}
\rho_\phi &=&  - \frac{1}{16 \pi} \left( X -  2 \Lambda \right) \\
P_\phi &=& \frac{1}{16 \pi} \left[ \left( X -  2 \Lambda \right) + \frac{ \dot{X} }{ 3H }  \right] .
\end{eqnarray}
The above expressions for the dark energy density and pressure are valid in the space of solutions with vanishing scalar current. 

\subsection{Application: power-law expanding universes}
\label{subsec:power_law}

We present an application of the recipe by associating the cosmological scenario described by
\begin{equation}
\label{eq:power_law_expansion}
a(t) = a_i \left( \frac{t}{ t_i } \right)^n
\end{equation}
where $a_i$ and $t_i$ are constants to a cubic Horndeski theory. This cosmological scenario encompasses the matter ($n = 2/3$) and radiation ($n = 1/2$) dominated scenarios in GR. The Hubble parameter following this is simply
\begin{equation}
H(t) = \frac{n}{t} .
\end{equation}
Substituting this into Eq. (\ref{eq:necessary_conserve_energy}) and solving the resulting differential equation leads to
\begin{equation}
\rho(t) = b t^{-3 n \left( 1 + w \right)}
\end{equation}
where $b$ is an integration constant. With Eq. (\ref{eq:necessary_X}), the kinetic density is then given by
\begin{equation}
X = 2 \Lambda - 6 \frac{n^2}{t^2} + 16 \pi b t^{-3n \left( 1 + w \right)} .
\end{equation}
Obviously, $X(t)$ can be numerically inverted to obtain $t(X)$. Thus, we can use Eq. (\ref{eq:current_constraint}) to build the only model function 
\begin{equation}
G_X = - \frac{1}{3\sqrt{2X}} \frac{1}{ H \left( t \left( X  \right) \right) }
\end{equation}
compatible with the cosmological scenario given by Eq. (\ref{eq:power_law_expansion}). Moreover, the inversion can performed analytically for $3n \left( 1 + w \right) = -1, 0, 1, 2, 3, 4$. GR's cosmological picture is part of these analytically solvable cases as we now explicitly show. 

For GR's matter-dominated scenario, $w = 0$ and $n = 2/3$, the kinetic density becomes
\begin{equation}
X = 2 \Lambda - \left( \frac{8}{3} - 16 \pi b \right) \frac{1}{t^2} .
\end{equation}
This is easy to invert for $t(X)$ and so the resulting model function is given by
\begin{equation}
G_X = - \frac{ \sqrt{-1 + 6 \pi b} }{ \sqrt{3X} \sqrt{ X - 2 \Lambda } }
\end{equation}
or
\begin{equation}
G(X) = - \frac{2}{\sqrt{3}} \sqrt{-1 + 6 \pi  b} \ln \left(\sqrt{X - 2 \Lambda}+\sqrt{X}\right) + C
\end{equation}
where $C$ is an integration constant. We see here that $b$ must satisfy the constraint $b > 1/6 \pi$ for the model function to be real. For GR's radiation-dominated case, $w = 1/3$ and $n = 1/2$, the kinetic density becomes
\begin{equation}
X = 2 \Lambda - \left( \frac{3}{2} - 16 \pi b \right) \frac{1}{t^2} .
\end{equation}
This almost looks like the kinetic density for the previous GR matter-dominated scenario except that the term $3/2$ in the parenthesis appears instead of $8/3$. The model is as easy to construct as well and the result is
\begin{equation}
G_X = - \frac{ \sqrt{-3 + 32 \pi b} }{ 3 \sqrt{X} \sqrt{ X - 2 \lambda } }
\end{equation}
or
\begin{equation}
G(X) = - \frac{2}{3} \sqrt{-3 + 32 \pi  b} \ln \left(\sqrt{X - 2 \Lambda}+\sqrt{X}\right) + C
\end{equation}
where $C$ is an integration constant. In this case, $b$ has to satisfy $b > 3 / 32 \pi$ for the model to be real. These explicit examples reveal the cubic Horndeski theories that can equally-well describe power-law expanding universes in GR.

We summarize in Table~\ref{tab:summary_power_law} the other cases for which it is still practical to write down the model function for the generic power-law expanding universe scenerio described by Eq.~(\ref{eq:power_law_expansion}).
\begin{table}[h!]
\centering
\caption{Some cubic model functions for Eq. (\ref{eq:power_law_expansion}).}
\begin{tabular}{|c|c|}
\hline \hline
$\ \ \ \ k = 3n \left( 1 + w \right) \ \ \ \ $ $\phantom{\frac{\frac{1}{1}}{\frac{1}{1}}}$ & \ \ \ \ model function $G_X$ \ \ \ \ \\ \hline \hline
0 $\phantom{\frac{\frac{1}{1}}{\frac{1}{1}}}$ & $- 1 /  \sqrt{ 3X } \sqrt{ 16 \pi  b- (X - 2 \Lambda ) } $ \\ \hline
1 $\phantom{\frac{\frac{1}{1}}{\frac{1}{1}}}$ & $ - \left( 8 \pi b + \sqrt{ - 6 n^2 \left( X - 2 \Lambda \right) + 64 \pi^2 b^2 } \right) / 3 \sqrt{2} n \sqrt{X} \left( X - 2 \Lambda \right)$   \\ \hline
2 $\phantom{\frac{\frac{1}{1}}{\frac{1}{1}}}$ & $- \sqrt{ - 3 n^2 + 8 \pi b } / 3 n \sqrt{X} \sqrt{ X - 2 \Lambda } $ \\ \hline
4 $\phantom{\frac{\frac{1}{1}}{\frac{1}{1}}}$ & $ - \sqrt{ - 3n^2 + \sqrt{9 n^4 + 16\pi b ( X - 2 \Lambda )}  } / \left( 3 \sqrt{2} n \sqrt{X} \sqrt{X - 2 \Lambda } \right) $ \\ \hline \hline
\end{tabular}
\label{tab:summary_power_law}
\end{table}
GR's matter and radiation dominated scenarios belong to the case $k = 2$. The model function for cases $k = -1$ and $k = 3$ can be analytically written down; however, the expressions are unwieldy and uninformative so we did not include them in Table~\ref{tab:summary_power_law}. The case $k = 3n \left( 1 + w \right) = 0$ is particularly interesting since the exponent $n$ quantifying the rate of the expansion does not appear in the model function defining the cubic Horndeski theory. Leaving out the case of $n=0$ (non-expanding universe), the more interesting case here seems to be $w = -1$ which is when the matter perfect fluid is dominated by a contribution from a cosmological constant. The case $k = 0$ shows that it is possible for a cosmological constant perfect fluid to drive an expansion such as Eq.~(\ref{eq:power_law_expansion}) that we typically associate with other fluid components, e.g. matter or radiation, in GR. The exponent $n$ for the $k = 0$ case is not fixed by the theory or the model. 

\subsection{Application: cubic Horndeski model for $\Lambda$CDM's Hubble evolution}
\label{subsec:cubic_lcdm}

\begin{figure}[h!]
	\subfigure[ ]{
		\includegraphics[width = 0.48 \textwidth]{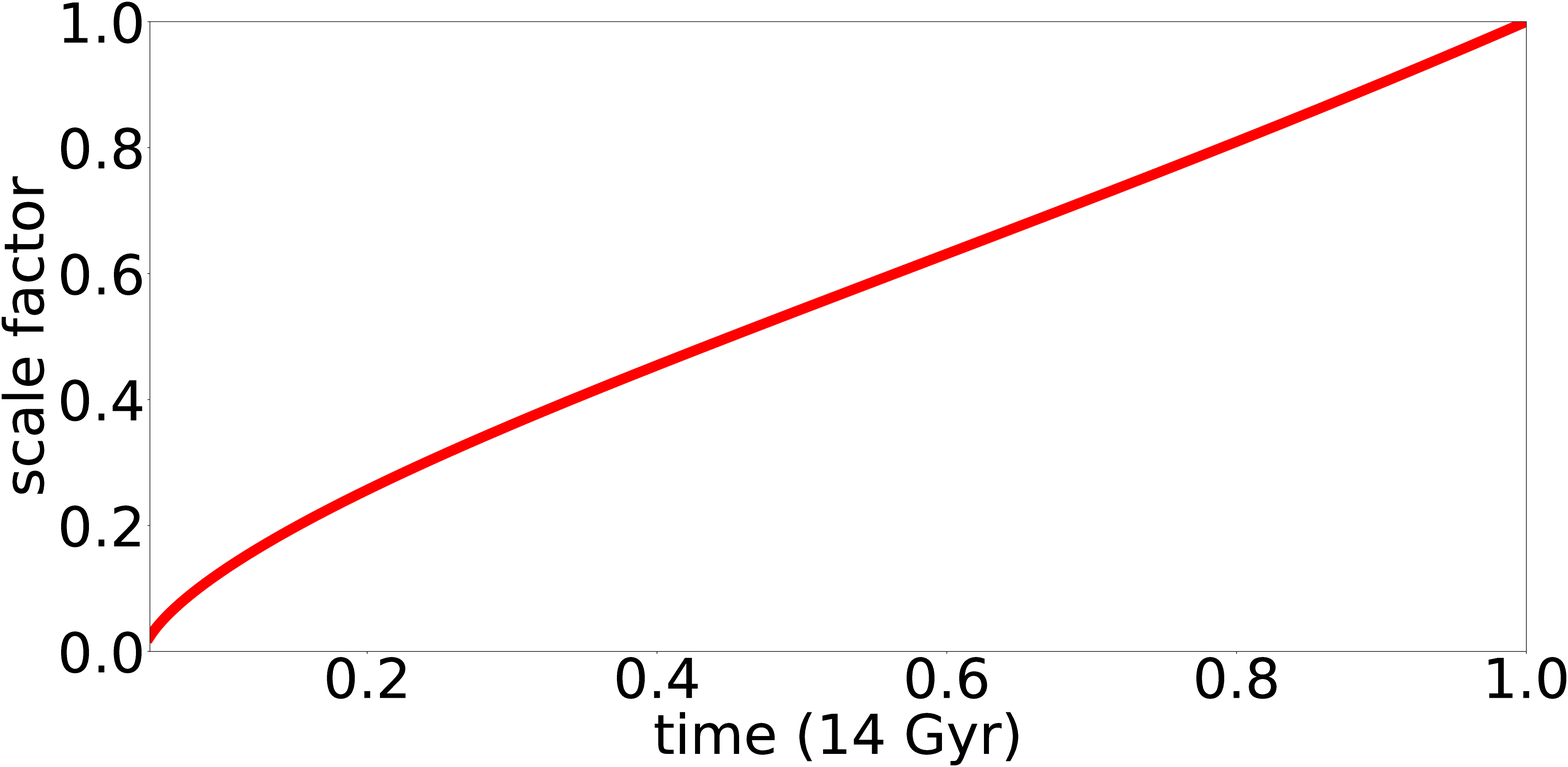}
		}
	\subfigure[ ]{
		\includegraphics[width = 0.48 \textwidth]{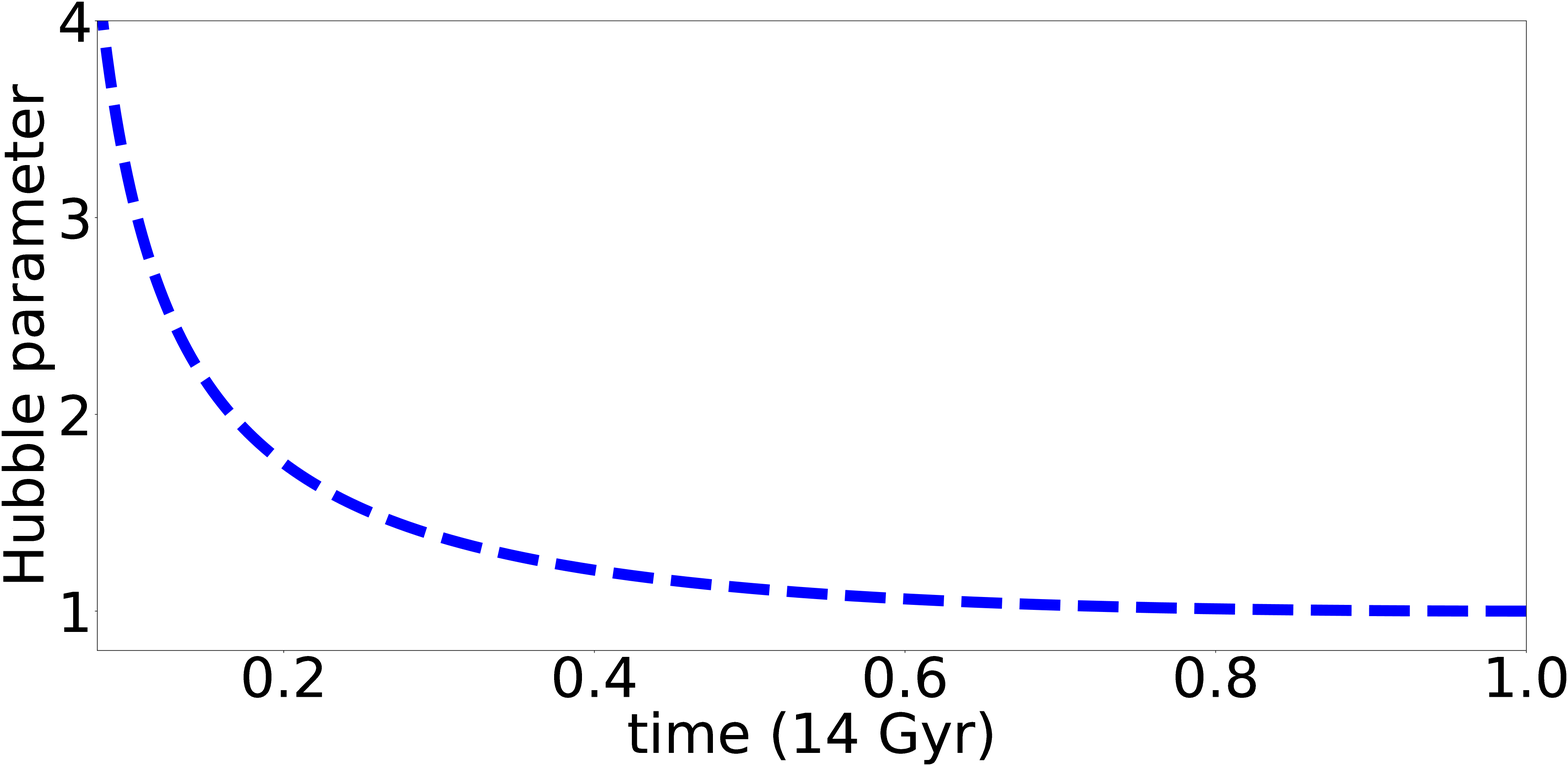}
		}
\caption{Evolution of $\Lambda$CDM scale factor and Hubble parameter.}
\label{fig:lcdm_universe}
\end{figure}

\begin{figure}[h!]
\center
	\subfigure[ ]{
		\includegraphics[width = 0.48 \textwidth]{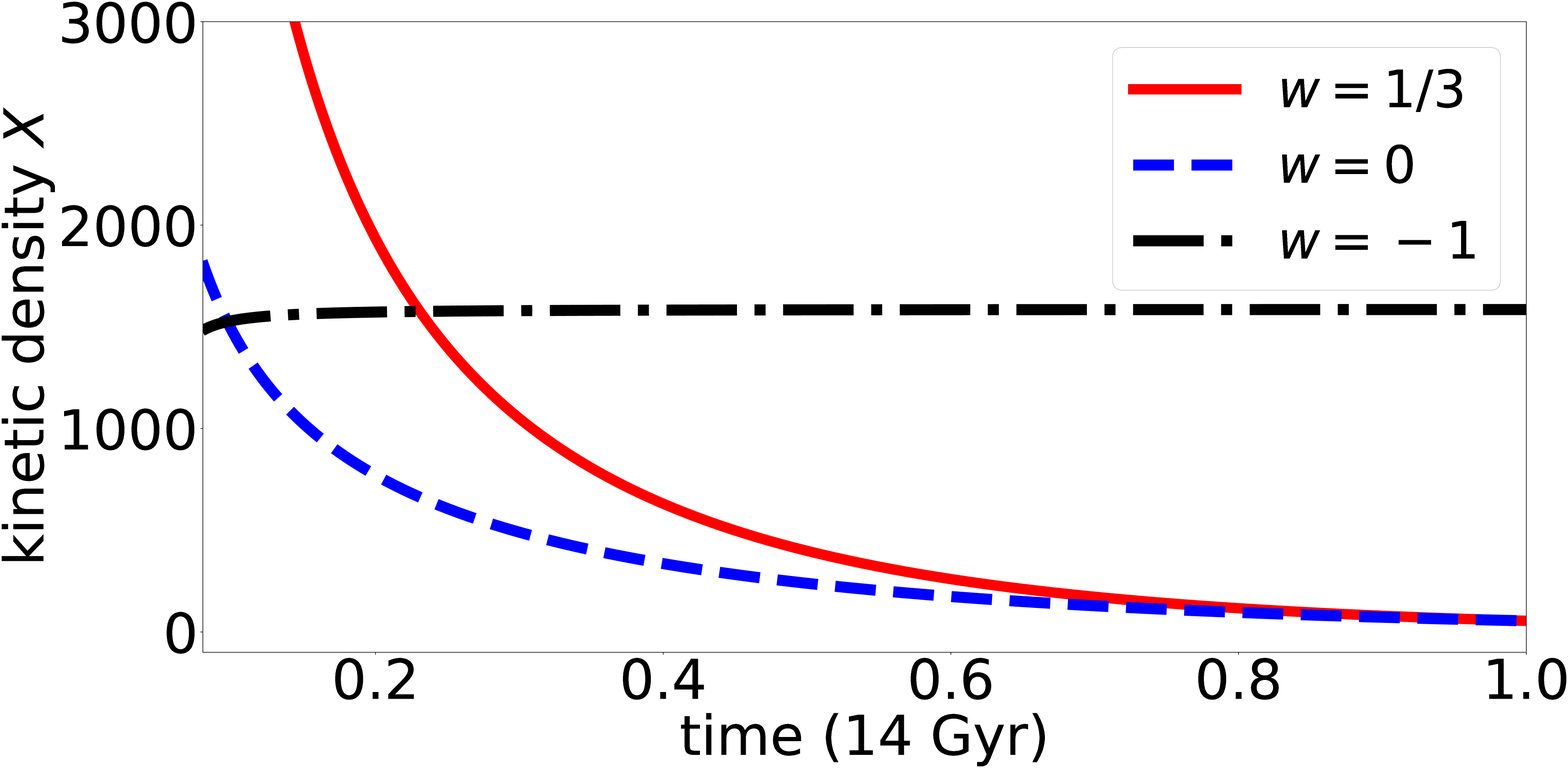}
		}
	\subfigure[ ]{
		\includegraphics[width = 0.48 \textwidth]{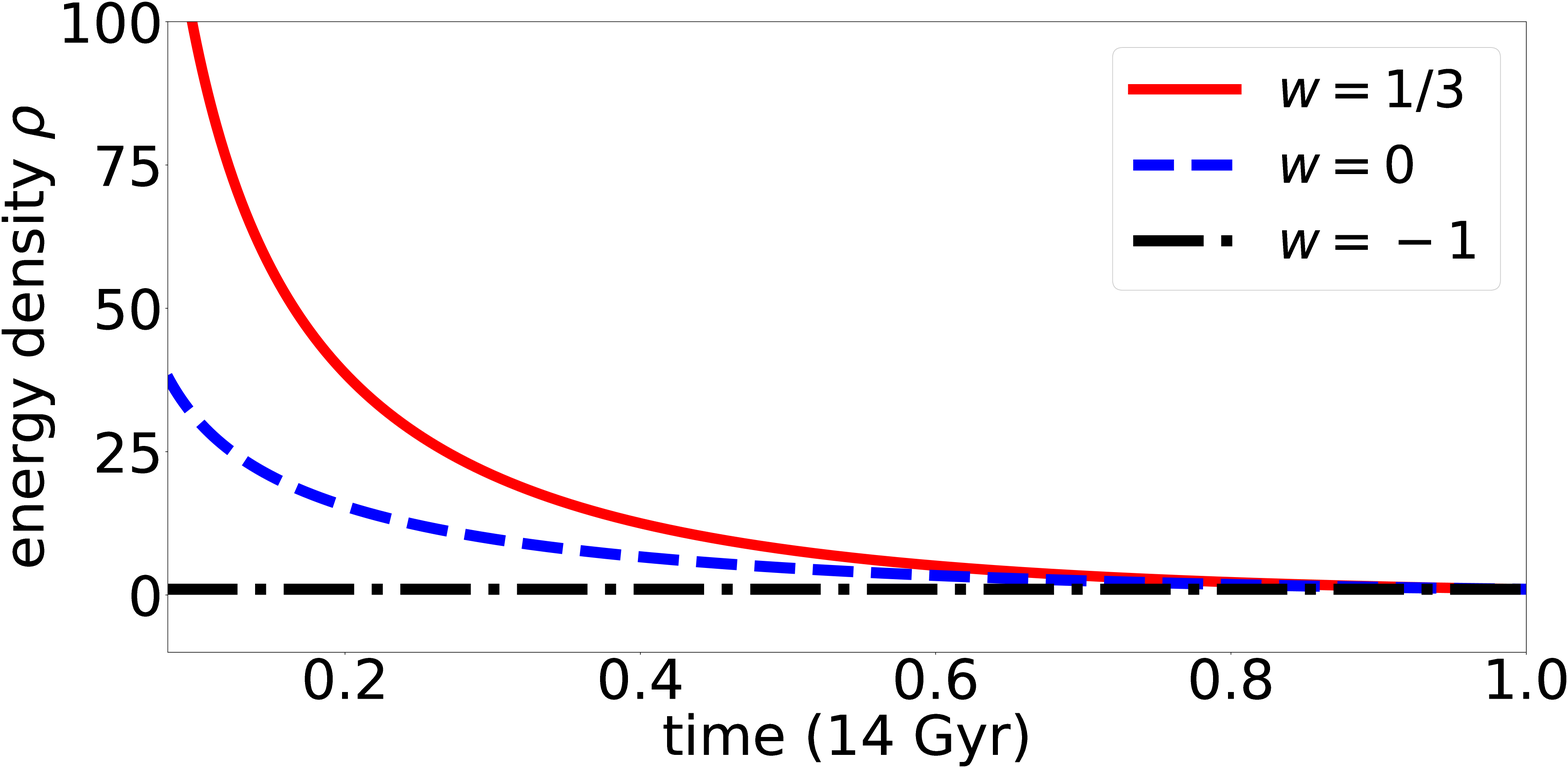}
		}
	\subfigure[ ]{
		\includegraphics[width = 0.48 \textwidth]{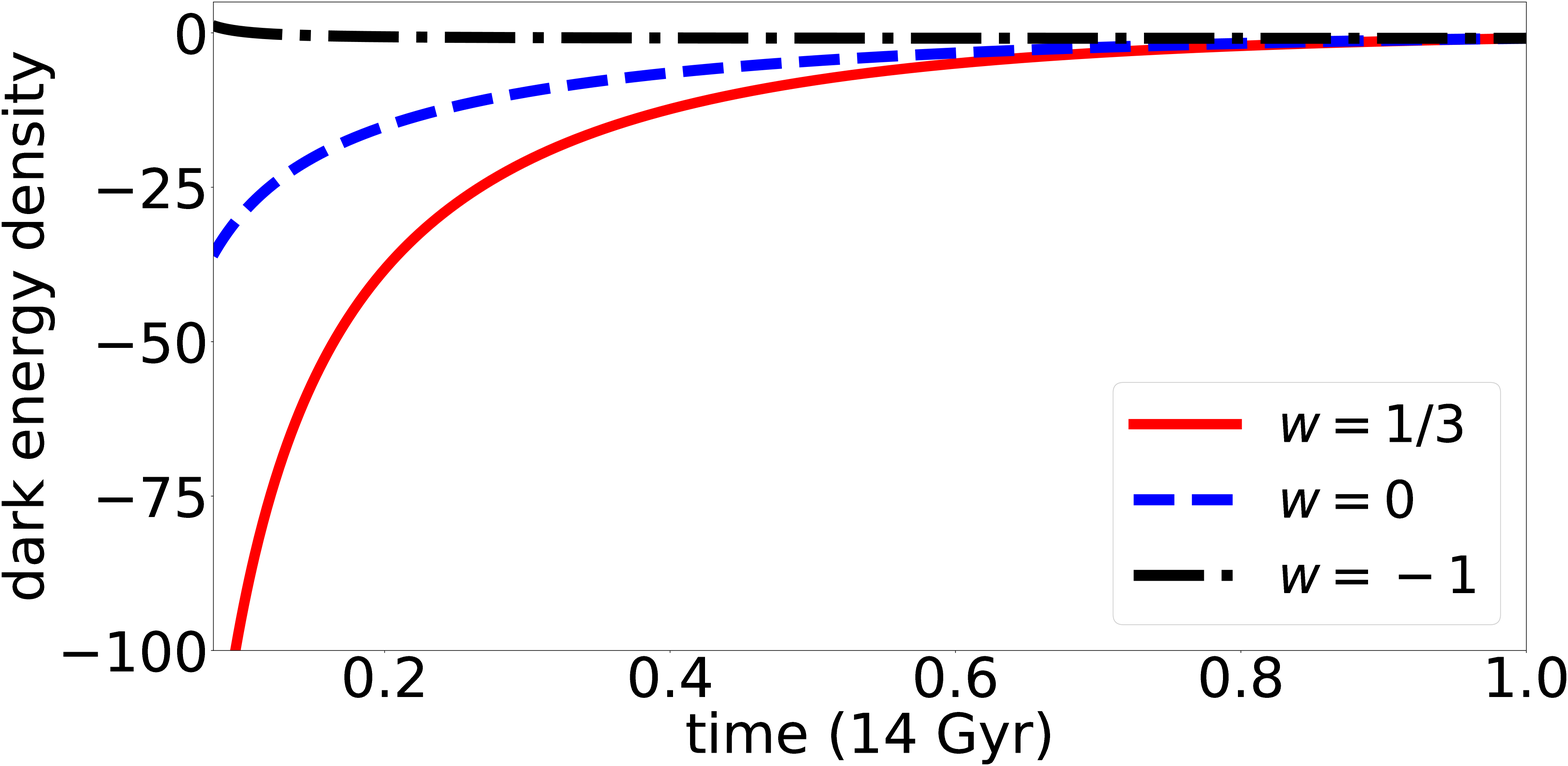}
		}
	\subfigure[ ]{
		\includegraphics[width = 0.48 \textwidth]{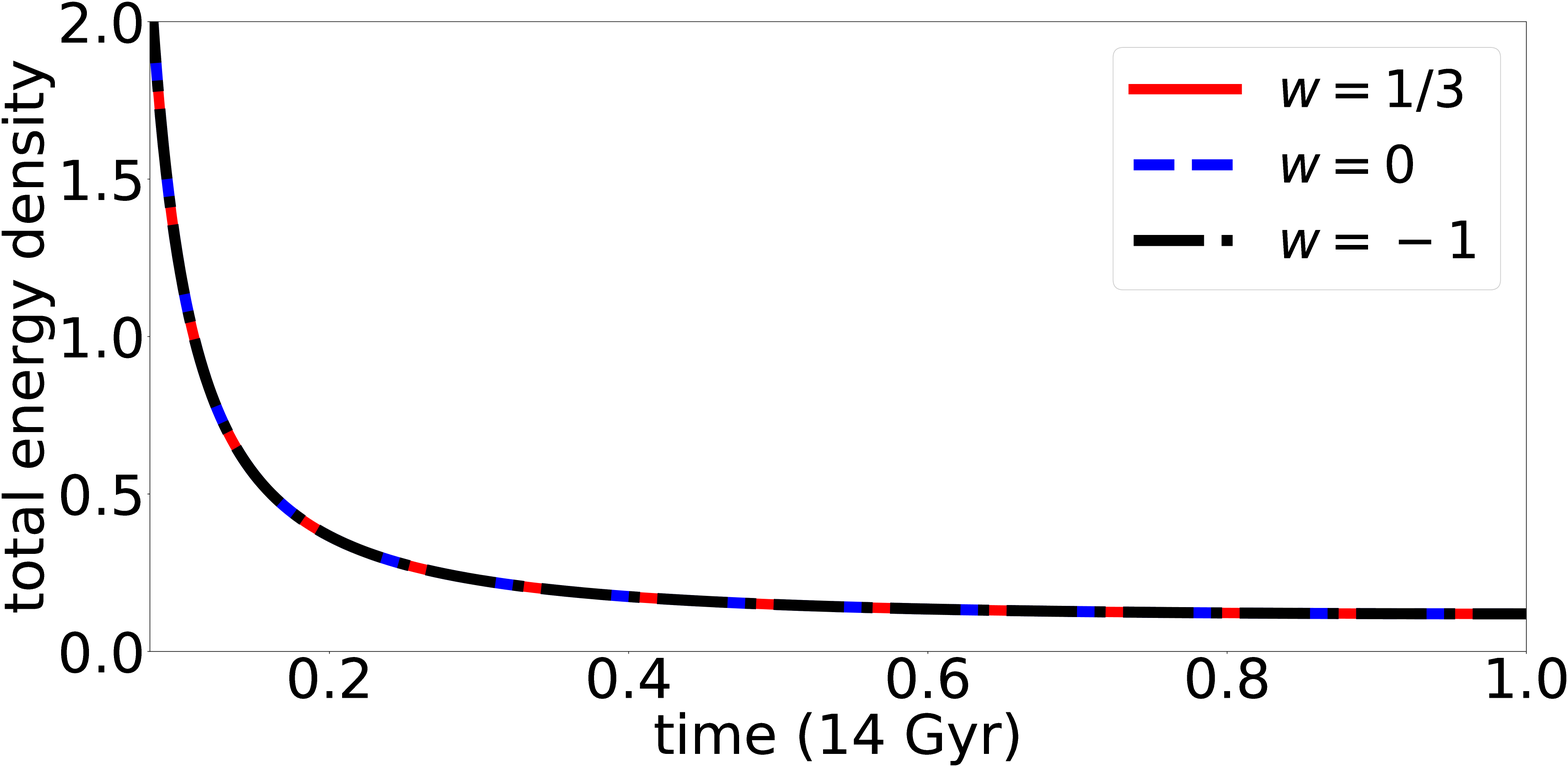}
		}
\caption{(a) Scalar field kinetic density $X$, (b) perfect fluid energy density $\rho$, (c) dark energy density, and (d) total energy density for baryonic matter ($w = 0$), radiative ($w = 1/3$), and vacuum ($w = -1$) perfect fluids in cubic Horndeski theory for the $\Lambda$CDM scale factor evolution displayed in Fig.~\ref{fig:lcdm_universe}.}
\label{fig:fit}
\end{figure}

\begin{figure}[h!]
\center
	\subfigure[ ]{
		\includegraphics[width = 0.48 \textwidth]{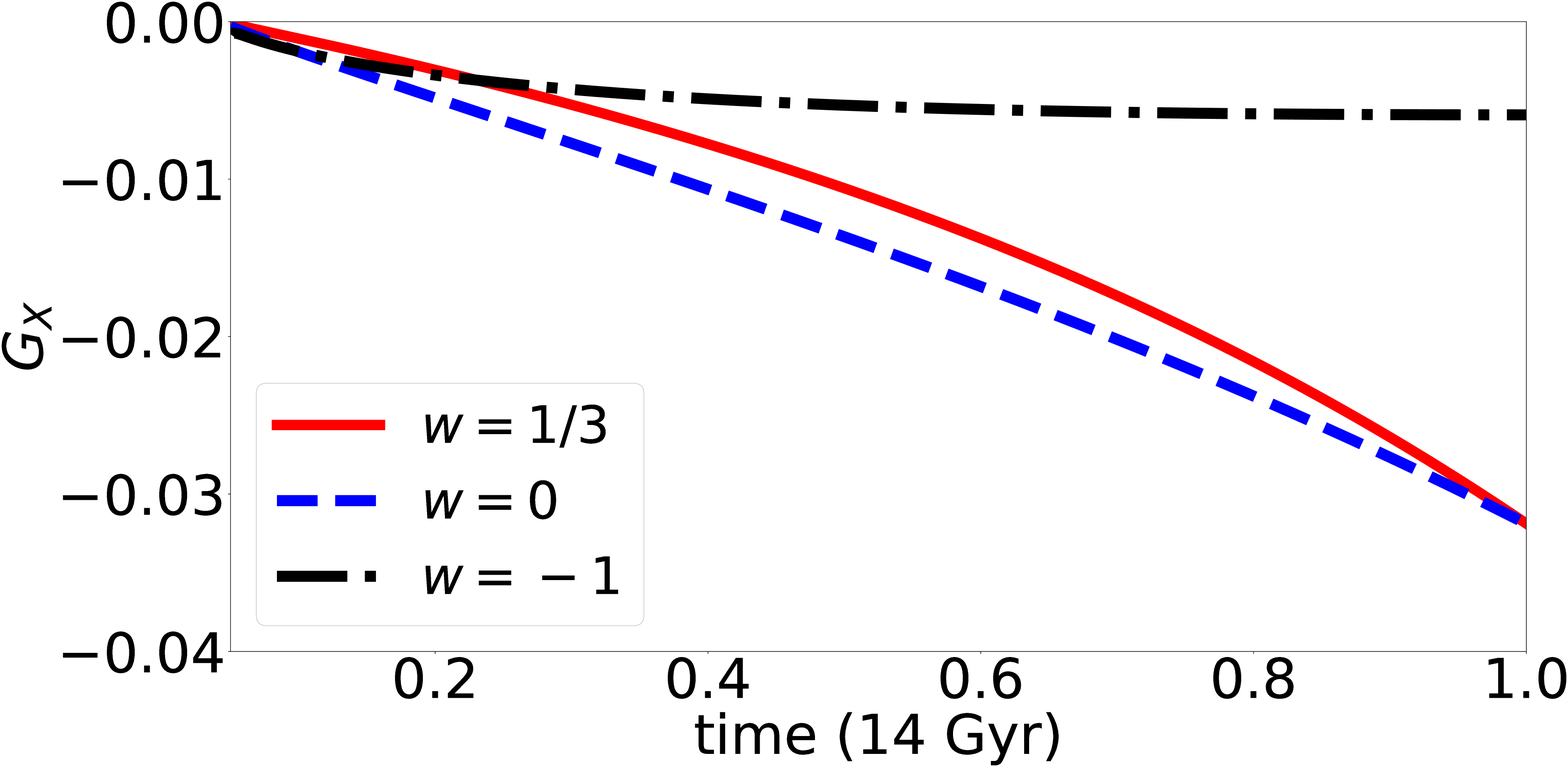}
		}
	\subfigure[ ]{
		\includegraphics[width = 0.48 \textwidth]{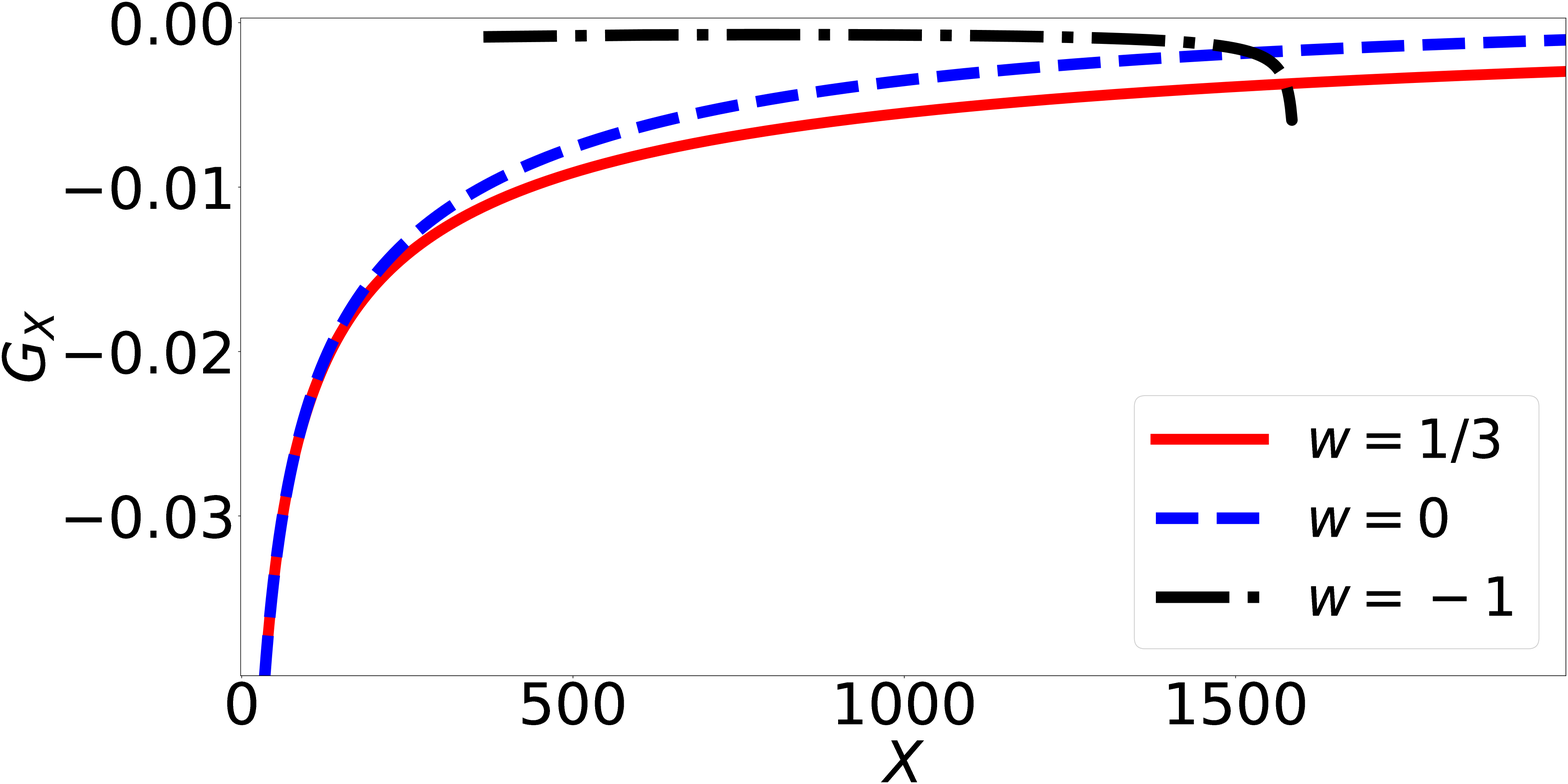}
		}
\caption{Cubic Horndeski models for matter ($w = 0$), radiation ($w = 1/3$), and vacuum ($w = -1$) equations-of-state in cubic Horndeski for the $\Lambda$CDM scale factor evolution displayed in Fig.~\ref{fig:lcdm_universe}.}
\label{fig:cubic_model_lcdm}
\end{figure}

\begin{figure}[h!]
	\subfigure[ ]{
		\includegraphics[width = 0.48 \textwidth]{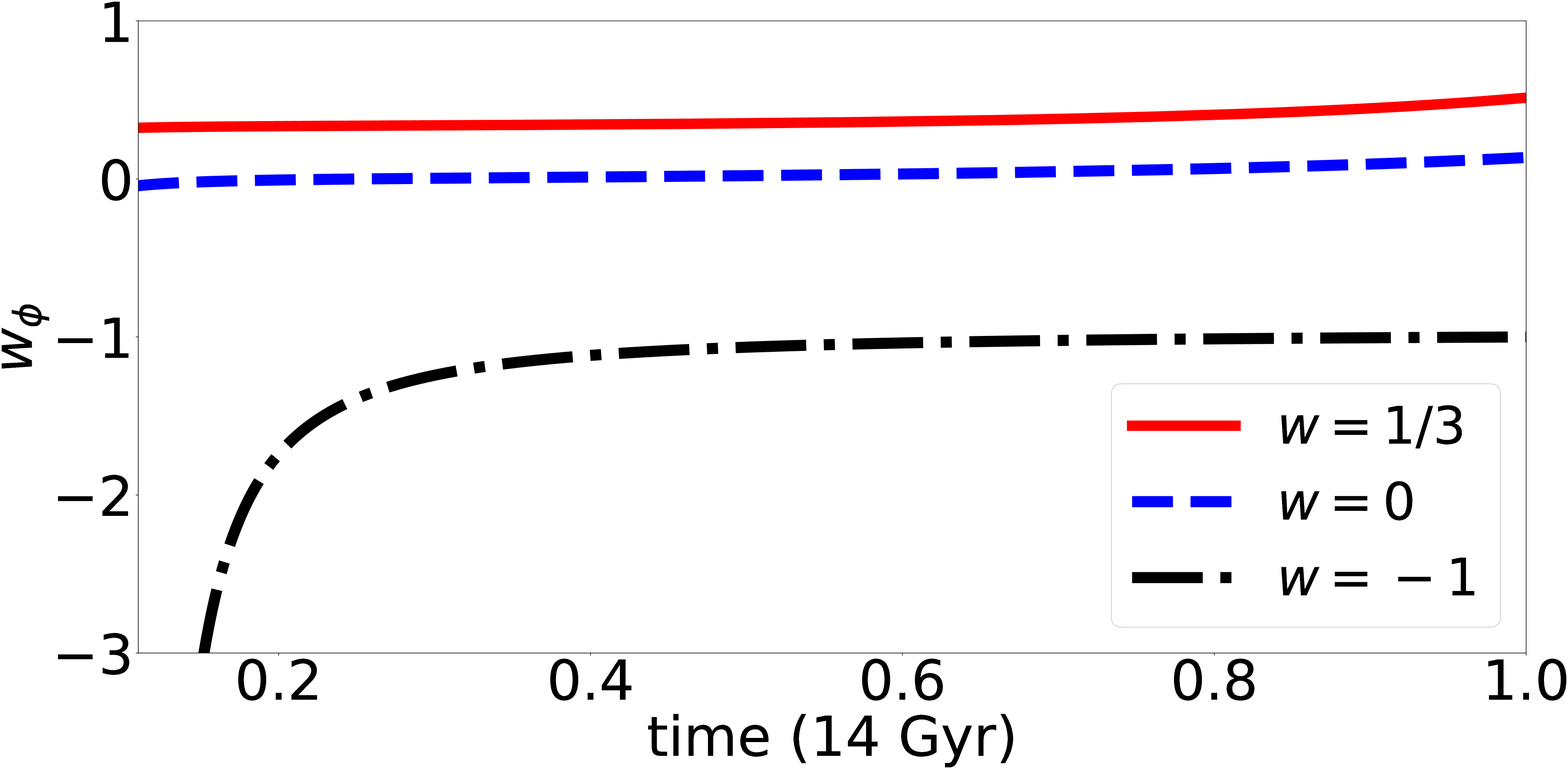}
		}
	\subfigure[ ]{
		\includegraphics[width = 0.48 \textwidth]{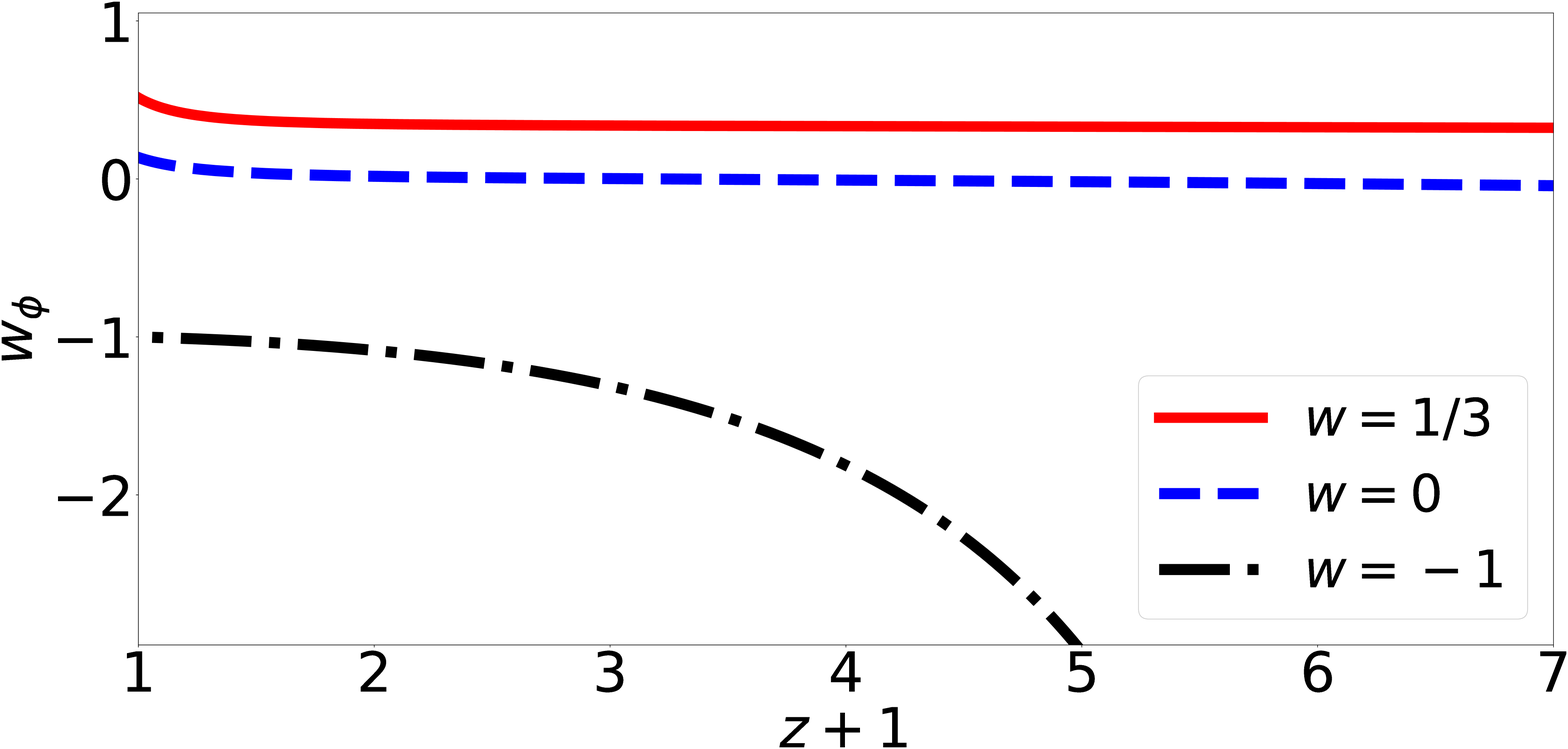}
		}
\caption{Dark energy equation-of-state, as a function of the comoving time and $z + 1$ where $z = 1/a - 1$ is the redshift, for the cubic Horndeski theories in Fig.~\ref{fig:cubic_model_lcdm}.}
\label{fig:w_de}
\end{figure}

For our next application of the recipe, we show that we can assign $\Lambda$CDM's predicted scale factor evolution into a cubic Horndeski model. Figure~$\ref{fig:lcdm_universe}$ shows $\Lambda$CDM's prediction of the scale factor and the Hubble parameter as a function of time. This is obtained by integrating GR's Friedmann equations with the matter and dark energy fractional densities determined by the Planck mission (see, for example, Table 2 of Ref. \cite{planck_2018}). In Fig.~\ref{fig:lcdm_universe}, the present value of the Hubble parameter is set to unity so that time is in units of $14$ billion years (Gyr).
Inserting either the scale factor or the Hubble parameter in Fig.~\ref{fig:lcdm_universe} as data for Eq.~(\ref{eq:necessary_X}), the integration of the necessary conditions for matter ($w = 0$) and radiation ($w = 1/3$) both with $\Lambda = 21$ and for vacuum ($w = -1$) with $\Lambda = 3081$ leads to the kinetic density, perfect fluid energy density, dark energy density, and total energy density displayed in Fig.~\ref{fig:fit}. The choice of $\Lambda$ is not arbitrary and was determined to be the minimum value to keep the kinetic density positive and associate a real cubic Horndeski theory. We note that in referring to the equation of state $w = -1$ we are referring as well to a cosmological constant term that enters the theory through the matter sector and is independent of the cosmological constant $\Lambda$ that is inherently included in the scalar field sector through the action given by Eq.~(\ref{eq:theory}).
The results show that, for any component of the perfect fluid, the energy densities generally drop in time which is expected given the adiabatic expansion of the universe. Even in cubic shift-symmetric Horndeski theory with the vanishing scalar current, the expansion has to be adiabatic since we obtained Eq. (\ref{eq:necessary_conserve_energy}) as a necessary condition for the solutions. Thus, the energy needed to expand has to be taken away from the internal energy, in this case, in the form of the cosmic fluid. A possibly surprising feature of the solutions presented in figure \ref{fig:fit} is that the total energy density seems to be independent of the component of the cosmic fluid. The independence of the total energy density on the equation of state of the matter sector is a general result since one can work out from the necessary conditions the expression $H^2 = 8 \pi \rho_{\text{total}} /3$ which is simply the first Friedmann equation. The corresponding model functions, or the cubic Horndeski theory, for the plots in figure \ref{fig:fit} are shown in figure \ref{fig:cubic_model_lcdm} for radiation, matter, and cosmological constant dominated perfect fluids.
These models represent the only three cubic Horndeski theories with vanishing scalar current that can describe $\Lambda$CDM's picture of the universe evolution. The dark energy equation of state for these theories are shown in Fig.~\ref{fig:w_de}.
It is shown that even though baryonic matter and radiation cosmic fluids can fit $\Lambda$CDM's Hubble parameter to a cubic Horndeski theory it is only the vacuum dominated cosmic fluid that can arrive at a dark energy equation of state $w_\phi$ that is close to $-1$ today. 

\subsection{Nondynamical dark energy}
\label{subsec:nondynamical_dark_energy}

As a final application of the recipe, we show that there is no cubic Horndeski theory with vanishing scalar current that can be assigned to a nondynamical dark energy equation of state $w_\phi = -1$. Also, apart from this case, we derive the only other scale factor which cannot be accomodated within the theory Eq. (\ref{eq:theory}) with vanishing scalar current.

For a general kinetic gravity braiding, the expression for the dark energy equation of state (Eq. (\ref{eq:w_de_kinetic_braiding})) can be reduced to Eq. (\ref{eq:w_de_H}) for the subset of the theory given by Eq. (\ref{eq:theory}) with vanishing scalar current. It is possible to show that the dark energy equation of state can also be written down as
\begin{equation}
\label{eq:w_de_X}
w_\phi = -1 - \frac{1}{3 H} \frac{\dot{X}}{X - 2 \Lambda} .
\end{equation}
A nondynamical $w_\phi = -1$ therefore implies that $\dot{X} = 0$ for all time. This implies that the kinetic density $X$ of the scalar field is a constant. Thus, we cannot invert the expression $X(t)$ and there is no way to make Eq. (\ref{eq:scalar_current_vanishes}) consistent with the rest of the field equations. This means that there is no cubic Horndeski theory with vanishing scalar current that can be associated with a nondynamical dark energy equation of state $w_\phi = -1$. In section \ref{subsec:inflation}, we have attempted to build a nondynamical dark energy equation of state by choosing the matter source as a cosmological constant. This attempt has only lead to freezing the scalar field, turning its energy to a constant, so that it becomes indistinguishable from the matter cosmological constant with the equation of state $w = -1$.

It can be insightful to compare Eq. (\ref{eq:w_de_X}) to the dark energy equation of state of quintessence
\begin{equation}
w_\phi^{\text{quintessence}} = - 1 + \frac{2 X}{X + V} .
\end{equation}
In quintessence, the dark energy field resembling a cosmological constant fluid in GR can be achieved with the slow-roll scalar fields, i.e. $X$ is small. For cubic shift-symmetric Horndeski theories with vanishing scalar current, the dark energy field instead has to satisfy the condition $\dot{X}$ is small to look like GR's cosmological constant fluid.

If the dark energy equation of state is therefore constrained to $w_\phi = -1$ by observation, the most that we can insist from a shift-symmetric cubic Horndeski theory with a vanishing scalar current is that the dark energy equation of state be equal to $w_\phi = -1$ today. It is, in fact, possible to formulate an alternative version of the recipe in which one supplies first the function $w(t)$ and then, from this input, develops the Hubble parameter, energy densities, and the cubic Horndeski theory. This alternative version begins with 
\begin{equation}
\label{eq:w_de_alternate_recipe}
w_\phi = -1 - \frac{1}{3 H} \frac{d}{dt} \ln \left( 8 \pi a^{-3\left( 1 + w \right)} - 3 H^2 \right) .
\end{equation}
The first step of the alternative recipe supplies $w_\phi(t)$ into Eq. (\ref{eq:w_de_alternate_recipe}) and then from this solves for the scale factor $a$ which is related to the perfect fluid density as $\rho = a^{-3\left(1 + w\right)}$. The subsequent steps are then the same, i.e. determine the kinetic density $X$ using Eq. (\ref{eq:necessary_X}) and then build the cubic Horndeski theory using Eq. (\ref{eq:scalar_current_vanishes}). 

Finally, we address the deeper question: ``what are the other Hubble parameters or scale factors that cannot be accomodated in cubic shift-symmetric Horndeski theory with the vanishing scalar current?" We answer this question by going back to the expression for $X$ given by Eq. (\ref{eq:necessary_X}). The general case that cannot be accomodated within the theory falls under the space of solutions described by the constraint
\begin{equation}
\label{eq:unaccomodated}
3 H^2 = 8 \pi \rho - \frac{Q}{2} 
\end{equation}
where $Q$ is a constant. This is simply the first Friedmann equation but with the dark energy density equal to a constant. Thus, in general, a nondynamical dark energy sector is incompatible with the theory. Within the constraint of Eq. (\ref{eq:unaccomodated}), it can be shown that the corresponding scale factor oscillates and is given by
\begin{equation}
\label{eq:unaccomodated_a}
a(t) = a_0 \cos^{2/3(1+w)} \left( \sqrt{\frac{3Q}{8}} \left(1 + w\right) \left(t - t_0\right) \right) .
\end{equation}
The evolution described by Eq. (\ref{eq:unaccomodated_a}) cannot be accomodated within the theory given by Eq. (\ref{eq:theory}) with vanishing scalar current.

\section{Conclusions}
\label{sec:conclusions}

The strong possibility that tensor perturbations propagate at the speed of light for all practical redshifts has dramatically reduced the landscape of viable alternative gravity theories, including Horndeski theory \cite{gw_170817_ligo, dark_energy_creminelli, dark_energy_ezquiaga, st_horndeski_cosmology_baker, st_horndeski_cosmology_sakstein, st_horndeski_cosmology_bettoni,st_horndeski_copeland2018}, for cosmological applications. A reaffirmation of luminally-propagating tensor perturbations from LISA - a future space-based gravitational wave observatory that works in the frequency band $10^{-4} - 10^{-1}$ Hz - will put to rest any further question about the speed of gravitational waves in the effective field theory of gravity \cite{dark_energy_derham}. Nonetheless, the space of viable Horndeski theories (and beyond) remain rather large, even with the imposition of numerous additional cosmological constraints  
\cite{st_horndeski_galileon_barreira_1, st_horndeski_galileon_barreira_2,st_horndeski_renk, st_horndeski_galileon_peirone,  horndeski_constraint_mancini2019, horndeski_constraint_noller2018, horndeski_constraint_komatsu2019,st_horndeski_copeland2018}. In this paper, we provide a tool for directly associating cubic shift-symmetric Horndeski theories with observations of the Hubble evolution. 

More, specifically we have presented a method (Section \ref{subsec:recipe}) that tailors a cubic shift-symmetric Horndeski theory (with vanishing scalar current) to a flat FRW spacetime with almost any given cosmological dynamics. This method extends ideas first applied in the search for hairy black hole solutions within the same Horndeski sector \cite{st_horndeski_hair_dressing_bernardo}. We have illustrated this method by obtaining exact analytical solutions and cubic Horndeski theories for inflation (Section \ref{subsec:inflation}) and power-law expanding universes (Section \ref{subsec:power_law}). We have also shown that $\Lambda$CDM's picture of the universe's evolution can be fitted to a cubic Horndeski theory with only a single component of cosmic fluid (Section \ref{subsec:cubic_lcdm}) and that it is not possible for a cubic shift-symmetric Horndeski theory with vanishing scalar current to accomodate a theory with a nondynamical dark energy equation of state $w_\phi = -1$ (Section \ref{subsec:nondynamical_dark_energy}). This and an oscillatory scale factor given by Eq. (\ref{eq:unaccomodated_a}) are the only cosmological dynamics that cannot be accomodated within the sector (Section \ref{subsec:nondynamical_dark_energy}).

A nice feature of our approach is that, because it is based on model-independent necessary conditions, it selects the \emph{only} compatible cubic Horndeski theory for a given Hubble parameter evolution. In contrast to the static and spherically-symmetric case \cite{st_horndeski_hair_dressing_bernardo}, Eq. (\ref{eq:necessary_conserve_energy}) is a linear differential equation for $a$ and $\rho$ so that the set $\{ a, \rho , \rho_\phi \}$ describing a cosmological scenario can be uniquely assigned to a cubic Horndeski theory $G_X$. 

The Horndeski sector we explore here is special in that only the free function $G_X$ appears in its field equations. This is the same sector that we previously investigated in seeking out hairy black holes in Horndeski theory \cite{st_horndeski_hair_dressing_bernardo}. An attempt to obtain similar model-independent necessary conditions for flat FRW solutions in other Horndeski sectors is more challenging. For instance, in shift-symmetric $k$-essence, $K$ and its first derivative $K_X$ enter the field equations through the scalar field energy density and pressure given by Eqs. (\ref{eq:dark_energy_density_general}) and (\ref{eq:dark_energy_pressure_general}). The situation is a lot more complicated for the quartic and quintic sectors of Horndeski theory since even second derivatives of the model functions appear in the expression for the scalar current. For these cases, the condition of vanishing scalar current appears too restrictive for a similar reduction of the equations. Our results are quite similar to what is already known with quintessence models. Though for our case, the required inversion of a cosmological dynamics to get the Horndeski model function $G_X$ (cf. quitessence's potential $V$) is entirely algebraic.

A natural direction to take from here is to square the cubic Horndeski theories presented in this paper with solar system screening tests \cite{screening_chakraborty, screening_brax, screening_schmidt, screening_davis,st_horndeski_vainshtein_dima, alternative_gravity_koyama, alternative_gravity_joyce} and cosmological observations \cite{eft_cosmology_piazza, eft_de_gleyzes2, eft_de_cremineli, eft_de_gubitosi, eft_de_gleyzes,st_horndeski_kennedy2017,
st_horndeski_kennedy2018,st_horndeski_kennedy2019}. The existence of Vainshtein screening in galileon gravity is well-known and the question of its applicability and other screening mechanisms to cubic Horndeki with vanishing scalar current is something that we intend to address in a future paper. One potentially useful framework for testing Horndeski theories against cosmological observables
has been laid down as the effective field theory of dark energy \cite{eft_cosmology_piazza, eft_de_gleyzes2, eft_de_cremineli, eft_de_gubitosi, eft_de_gleyzes,st_horndeski_kennedy2017,
st_horndeski_kennedy2018,st_horndeski_kennedy2019}. In this approach, the background dynamics is reduced to a set of constant coefficients in front of the perturbations and the analysis is carried out agnostic of the background, i.e. an explicit scale factor, scalar field, and Horndeski model, for generality. It is interesting to see how cosmological perturbations grow on top of our background solutions. 
This is another venture that we intend to return to in the future.

\appendix

\section{Some geometric objects in FRW}
\label{sec:some_geometric_objects}

Some of the scalar field and curvature quantities that follow from Eqs. (\ref{eq:line_element}) and (\ref{eq:scalar_field_spatially_uniform}) are
\begin{eqnarray}
X &=& \frac{ \dot{\phi}^2 }{2} \\
\partial_\alpha \phi &=& \delta_{\alpha}^0 \dot{ \phi } \\
\partial^\alpha \phi &=& - \delta^\alpha_0 \dot{ \phi } \\
\nabla_\alpha \nabla_\beta \phi &=& \delta_\alpha^0 \delta_\beta^0 \ddot{\phi} - \Gamma^0_{\alpha \beta} \dot{ \phi } \\
\Box \phi &=& -\ddot{\phi} - 3 \frac{ \dot{a} }{a} \dot{\phi} .
\end{eqnarray}
The nonzero Christoffel connections are
\begin{eqnarray}
\Gamma^x_{tx} &=& \Gamma^y_{ty} = \Gamma^z_{tz} = \frac{\dot{a}}{a} \\
\Gamma^t_{xx} &=& \Gamma^t_{yy} = \Gamma^t_{zz} = a \dot{ a }
\end{eqnarray}
and the nonzero Einstein tensor components are
\begin{eqnarray}
G_{xx} &=& G_{yy} = G_{zz} = -\dot{a}^2 - 2 a \ddot{a} \\
G_{tt} &=& 3 \left( \frac{\dot{a}}{a} \right)^2 .
\end{eqnarray}

\acknowledgments
The authors are grateful to Shuang-Yong Zhou and Joseph Bunao for constructive criticism of the preliminary version of the manuscript. This research is supported by the University of the Philippines OVPAA through Grant No.~OVPAA-BPhD-2016-13.





\bibliographystyle{JHEP}

\providecommand{\href}[2]{#2}\begingroup\raggedright\endgroup

\end{document}